\documentclass[%
 ,twocolumn%
 ,secnumarabic%
 ,showpacs,showkeys
,amssymb, amsmath,nobibnotes, aps, prd]{revtex4}

\usepackage{graphicx}
\usepackage{dcolumn}
\usepackage{bm}

\usepackage{epsfig} 

\newcommand{\be}{\begin{equation}}
\newcommand{\ee}{\end{equation}}
\newcommand{\bea}{\begin{eqnarray}}
\newcommand{\eea}{\end{eqnarray}}
\newcommand{\bmu}{\begin{multline}}
\newcommand{\emu}{\end{multline}}
\newcommand{\nn}{\nonumber}

\def\MES{{mass eigenstate}}

\def\msun{{\,M_\odot}}

\def\simlt{\lower.5ex\hbox{$\; \buildrel < \over \sim \;$}}
\def\simgt{\lower.5ex\hbox{$\; \buildrel > \over \sim \;$}}

\def\cm{{\rm\,cm}}
\def\km{{\rm\,km}}

\def\kpc{{\rm\,kpc}}

\def\gcm3{{\rm\,g\,cm^{-3}}}
\def\ncm3{{\rm\,cm^{-3}}}

\def\>{$>$}
\def\<{$<$}

\begin{document}
\title{Neutrino Interferometry In Curved Spacetime}
\author{Roland M. Crocker}
\email{r.crocker@physics.unimelb.edu.au}
\affiliation{
School of Physics, 
The University of Melbourne, 3010, Australia}
\author{Carlo Giunti}
\email{giunti@to.infn.it}
\homepage{http://www.to.infn.it/~giunti}
\affiliation{INFN, Sezione di Torino, and Dipartimento di Fisica Teorica, 
Universit\`a 
di Torino,
Via P. Giuria 1, I--10125 Torino, Italy}
\author{Daniel J. Mortlock}
\email{mortlock@ast.cam.ac.uk}
\affiliation{Institute of Astronomy, 
Madingley Road, Cambridge CB3 0HA, United Kingdom}
\date{\today}            
\begin{abstract}
Gravitational lensing introduces the possibility of multiple (macroscopic) paths
from an astrophysical neutrino source to a detector. Such a multiplicity 
of paths can allow for 
quantum mechanical interference  to take place that is qualitatively 
different to
neutrino oscillations in flat space. After an illustrative example clarifying 
some
under-appreciated subtleties of the phase calculation, we derive the form of 
the quantum
mechanical phase for a neutrino \MES \ propagating non-radially through a 
Schwarzschild metric.
We subsequently determine the form of the interference pattern seen at a 
detector. 
We show that 
the neutrino signal from a supernova
could
exhibit the interference effects we discuss 
{\it were} it lensed by an object in a suitable mass range. 
We finally conclude, however, that -- given current
neutrino detector technology -- the probability of such lensing
occurring for a (neutrino-detectable) supernova is tiny in the immediate future.
\end{abstract}
\pacs{14.60.Pq, 95.30.Sf, 98.62.Sb}
\keywords{Neutrino Oscillations, Gravitational Lensing}
\maketitle

\section{Introduction}


Spacetime curvature allows, in general, for
there to be more than one macroscopic path from a particle source to a detector.
This means that
there is a  quantum mechanical
interference phenomenon that may occur
-- at least in principle --
with gravitationally-lensed, astrophysical neutrinos that is
qualitatively different from `conventional'
neutrino oscillation. 
The possibility for this
different type of
interference arises because -- with, generically, 
each path from source to detector having a different
length -- a phase difference may develop at the detector due to affine {\it path}
difference(s). This is to be contrasted with flat spacetime 
 neutrino oscillations which arise because different mass eigenstates
 generically have different {\it phase} velocities. 
One might expect, in fact, that 
gravitationally-induced neutrino interference (`GINI')
exhibit a phenomenology partially analogous  to 
that produced by a Young's double slit experiment, viz., regular patterns of 
maxima and minima 
 across a detected energy spectrum.
As we show below, for ultra-relativistic neutrinos,
each maximum and minimum at some particular energy is characterised by, respectively, 
enhancement and depletion
of {\it all} 
neutrino species (not relative depletion of one species with respect to another which
characterizes flat space neutrino oscillations).

Below we shall provide the theoretical
underpinning to all the contentions made above. 
We also sketch a proof-of-principle that this interference effect
could actually be seen in the neutrinos detected from a  supernova
{\it given} a suitable lens.
There are
other situations where the GINI
effect might, in principle, also be evident. Reluctantly, however, we conclude that
pragmatic considerations
mean that GINI effects will be very difficult to see in these cases.

\section{Survey}

Particle interferometry experiments enjoy a venerable lineage and  -- apart from their
intrinsic interest -- have often found utility in the measurement of intrinsically small
quantities. The idea that the effects
of {\it gravity} -- the epitome of weakness as far as particle physics is concerned -- 
on the phase of particles might become manifest in
interferometry dates to the seminal, theoretical
 work of Overhauser and Colella \citep{Overhauser1974}. 
It was these researchers  themselves, together
with Werner \cite{Colella1975}, who
were the first to experimentally confirm the effect they were
predicting (in what has come to be labeled a COW experiment after the initials of these
researchers: see Ref.\cite{Greenberger1979}
for a review).

Another interesting idea involving gravitational effects on interferometry of
neutral particles
-- though,
 to the authors' knowledge, without yet having received experimental confirmation --
is the idea that gravitational micro-lensing of {\it light} might realise a {\it de facto}
Young's double slit arrangement. There is an extensive literature devoted to this idea
(see 
Refs.~\cite{Mandzhos1981,Ohanian1983,Schneider1985,Deguchi1986a,Deguchi1986b,
Peterson1991,Gould1992,Stanek1993,Ulmer1995}), 
which has been
labeled `femtolensing' because of the natural angular scales involved
for cosmologically-distant sources and lenses \citep{Gould1992}.
Femto-lensing is somewhat more closely analogous
to the idea we present (indeed, as we show below, 
the analogy becomes exact  in the
massless neutrino limit) than COW-type experiments.
 This is   because in femtolensing gravity
not only affects the phase of the propagating photons, but
is also
itself responsible for 
the `bending' of these particles so that diverging
particle beams (or, more precisely, wave packets) can be brought back together to interfere.
Furthermore, while the interfering particles are relativistic 
in both the femtolensing and GINI cases, they are non-relativistic in COW experiments.

As far as sources go,
light from GRBs 
has received particular
attention in the context of femtolensing \citep{Gould1992,Stanek1993,Ulmer1995}.
While we would, of course, also require astrophysical objects as the sources for
a GINI `experiment', the  sources  best able to offer a chance for the detection
of this effect are probably supernovae.
A Galactic supernova would generate excellent statistics in
existing solar (and other) neutrino detectors (thousands of events -- see below).
And with a  much larger, generation of neutrino detectors on the drawing
board -- some having as one of their chief design goals the detection of neutrinos
from supernovae occuring almost anywhere in our Local Group -- prospects
for the detection of the effect we predict can only improve with time
\footnote{Note here 
that while there is quite some similarity between GINI and femtolensing effects,
there are at least two effects that might mean that
neutrino interference  be, at a pragmatic level,
intrinsically more observable than femtolensing:
(i) the typical length scale for the impact paramter
in gravitational lensing is given by the Einstein radius of the
lens. It may happen that 
the source-lens-observer geometry means
that the Einstein radius of a lens is actually `inside'
the lens body. There will be many situations, then, where
the lensing object is optically thick to photons 
at the Einstein radius but transparent to neutrinos, meaning that
interference effects are, in principle, observable in the former
situation but not the latter.
(ii) interference
effects can only show up 
when different lensed images are unresolved (i.e., 
one's apparatus must {\it not} be able to determine
which photon -- or neutrino -- belongs to which image).
But, because of the
very different, {\it intrinsic} angular resolutions of the microscopic
processes
 involved in neutrino and photon detection,
a clearly-resolved, astrophysical light source  may well be,  an 
{\it un}resolved source as far as
neutrinos are concerned.}. 

By way of a pedagogical detour, 
please note the following: we believe the `time-delay' nomenclature
is misleading in the context of either femtolensing or GINI effects. It is
much better, we contend, to think in terms of path difference(s). The idea of
a well-defined time-delay belongs to classical physics. The time delay is -- in the
frame of some observer -- the time elapsed between the arrival of two signals.
These should have their origins in the `same' (macroscopic) 
event at a source, but then travel down
different classical geodesics from source to detector. Now, from the viewpoint
of quantum mechanics, 
there is a limit in which the classical description just given makes sense and 
is useful. This limit is that
in which the size of the wavepackets describing the
signaling particles is small in comparison to the affine 
path length differences between the different classical trajectories under consideration.
This limit will usually be satisfied in observationally-interesting
cases of gravitational lensing. But this limit
must {\it not} be satisfied if femtolensing or GINI effects are to be observed. Indeed,
we require the opposite situation to pertain, viz, an affine path length difference
of the order or smaller that the wavepacket size. This is required
so that wave packets 
created in the same (microscopic) event can overlap at the detector --
with interference effects being the result. In this sense, there is {\it no} 
time delay because
the wavepackets have to be overlapping at the detector position at the same (observer) time,
i.e., overlap must be satisfied at the spacetime location of the observation event.
Note further that, to paraphrase Dirac, each photon -- or neutrino -- only interferes
with itself. So it is the wavepacket of the single particle that results from a
single (microscopic) event -- like the decay of an unstable parent particle -- 
that, in simple terms, splits to travel down all the classical geodesics from source
to detector, only to interfere when recombined there. The idea we are describing, then,
does not require some weird (and impossible) analog of a `neutrino laser'; it works
at the level of individual, particle wavepackets.

Another strand that will be peripherally 
drawn into this paper is the behavior -- at a classical level --
of neutrinos in a curved spacetime background (i.e., gravitational lensing of neutrinos
treated as ultra-relativistic, classical particles). This topic  
became of immediate interest with the detection of neutrinos from SN 1987A 
\cite{Hirata1987,Bionta1987}.
Timing information
from the {\it nearly} simultaneous detection of these neutrinos and the supernova's
photon signal and from the time and energy spread of the neutrino burst alone 
has been investigated in many papers 
as an empirical limit on the  neutrino mass scale 
(see, e.g., Refs.~\citep{Bahcall1987, Arnett1987} and 
Refs.~\citep{Hillebrandt1989,Beacom2000,Bilenky2003} 
for reviews
and the seminal references concerning this idea)
and also
as a probe of the equivalence principle over
intergalactic distance scales \citep{Longo1988,Krauss1988}\footnote{Data from
SN 1987A neutrinos have been used to contrain other neutrino 
properties including neutrino mixing
and mass hierarchy, neutrino lifetime, and neutrino magnetic moment: 
see Ref.~\citep{Bilenky2003}
for a review.}.
More speculatively but germane to this work,
 the apparently bi-modal
distribution of  SN 1987A neutrinos observed by the then-operating 
Kamiokande solar neutrino observatory
was given an explanation in terms of an intervening gravitational lens (in the
$5 \times 10^5 \msun$ mass range:
\citep{Barrow1987}). Further, the idea that astrophysical neutrino `beams' might be
gravitationally focused by massive objects like the Sun has been investigated and
it has been found that such focusing can amplify an 
intrinsic neutrino signal by many orders of
magnitude (see \cite{Gerver1988} and \cite{Escribano2001} for more recent work).

An early and important work treating  the
 {\it quantum mechanical} aspects of 
neutrino propagation through a curved metric
 is that of  Brill and Wheeler \citep{Brill1957}.
Their work is particularly important for
its elucidation of
the formalism that allows one to treat
(massless) spinor fields under the influence of gravitational effects
(i.e., the extension of the Dirac equation to curved spacetime)

As presaged above, in this paper
we shall be particularly concerned with the {\it phase} of neutrinos
(more particularly, neutrino \MES s) in curved spacetime.
The seminal work treating the phase of quantum mechanical 
particles in curved spacetime is that of
Stodolsky \citep{Stodolsky1979}. In this work the author 
argued that the phase of a {\it spinless} particle in 
an arbitrary metric is identical with the particle's classical action (divided by $\hbar$).
Later work conducted on neutrino oscillations in curved spacetime 
\citep{Kojima1996,Fornengo1997,Bhattacharya1999},
has often -- though not always \citep{Cardall1997,Konno1998} 
-- implicitly assumed the correctness of  Stodolsky's 
contention (that the phase is given by 
the classical action) for spinor fields as well and taken this as its starting point. 
Somewhat ironically -- as we set out
in detail below -- recent researches \citep{Alsing2001} have revealed that 
the equality of classical action and phase holds for spin half particles, but {\it not} for
spin zero or one particles, or, at least not in an unqualified sense.
In any case, that Stodolsky's contention holds for spinors means a considerable
simplification for our calculations as we can avoid directly treating the covariant Dirac
equation.

A full review of the literature (see  
\cite{Ahluwalia1996,Kojima1996,Grossman1997,Cardall1997,Fornengo1997,Konno1998,
Ahluwalia1998,Bhattacharya1999,Alsing2001,Wudka2001,Zhang2001,Linet2002,Zhang2003})
on neutrino phase 
in 
the presence of gravity is beyond the scope of this work. Suffice it to say that
most work here to date has been concerned with the calculation of neutrino phase
in radial propagation of neutrinos through stationary, spherically symmetric, spacetimes.
There is active controversy in this context as to at what order in neutrino mass 
($m_\nu^2$ or $m_\nu^4$?)
gravitational corrections show up in the phase 
\citep{Ahluwalia1996,Konno1998,Bhattacharya1999,Wudka2001}. The answer to this
hangs critically on how energy and distance, in particular, are defined \citep{Wudka2001}.
We shall have to treat  such issues carefully, but all the subtleties of this 
debate need not particularly concern us. This is because we are {\it primarily} interested in
gravity not for its effect on phase {\it per se}, but for its ability to generate multiple
macroscopic paths from source to detector. And it is what might actually be
measured at the detector that concerns us. Detectors 
count the neutrinos -- 
registered in terms of flavor and (local) energy -- that interact within their volume.
 From these
one can infer interference patterns, but one does not, of course, have any direct
experimental access to the phase difference(s) (a point that does sometimes seem to
be forgotten). 

In regard to interference phenomenology, note 
the following:
whereas interference patterns 
with flat space neutrino oscillations take the form of
variations in neutrino flavor {\it ratios} across energy, with GINI, because there
will be constructive and destructive interference between the multiple allowed routes, 
 one (also) expects to see, in general,
maxima and minima (distributed across energy) 
in the counts of {\it all} neutrino flavors. 
These maxima and minima 
will be present irrespective of what measure of distance, say,
we settle on (though, of course, they may be undetectably small in amplitude -- 
but that is a separate issue). To put this in a different way, flat space neutrino
oscillations modify the {\it relative} abundances of neutrinos expressed as a function of
energy whereas GINI effects can modify {\it absolute} abundances.

Interestingly, 
of all the papers devoted to the topic of gravititationally-affected
neutrino phase, to the authors'
knowledge, only one \citep{Fornengo1997} has previously 
examined GINI, which, to reiterate, is
the idea that 
neutrino {\it oscillations} in the presence of
gravitational lensing -- or, to be strict, gravitational focusing -- might 
present interesting phenomenology. (This is the analog of the femto-lensing described
above that involved light.) To examine this idea, the authors of
 Ref.~\citep{Fornengo1997} were obliged to 
develop a formalism to deal with non-radial propagation of neutrinos
around a lensing mass, and we shall adopt much of this formalism in the current work.
 Unfortunately, Ref.~\citep{Fornengo1997}  contains
an incorrect result which it is one of the major aims of this paper to point out. Moreover,
other works which have considered gravitationally-affected neutrino
phase  contain results -- and commentary thereon -- 
which, if not strictly incorrect, can be misleading if one does not realise
the restricted nature of their tenability. In brief, most authors have failed to consider
the possibility of multiple paths. Any result which suggests the vanishing of neutrino phase
difference in the massless limit 
[see, e.g., Eq.~(13) of Ref.~\cite{Bhattacharya1999}
or 
Eq.~(4.7) of \cite{Konno1998}] should be interpretted with extreme caution
\footnote{As an example of this, take the contention on p.1483 of Alsing et al. 
in Ref.~\citep{Alsing2001}
that the phase of  photons propagating through a a Schwarzschild metric vanishes 
to lowest
order. This applies only to
radially-propagating photons and even then, should really be thought of as a 
statement about the {\it action}
along the classical, null-geodesic, rather than as a claim that the phase -- 
and therefore a potentially-measurable
phase difference -- actually is (close to) zero. Likewise, note that while the statement 
made in 
Ref.~\citep{Grossman1997} that 
`\dots if one compares two experimental setups with and without
gravity with the same curved distance in both cases there is no effect' 
is true,
it implicitly assumes that there is only one path from source to 
detector that 
need be considered
-- and this does not hold in general in the presence of gravity.}.

Essentially, the incorrect result in Ref.~\citep{Fornengo1997},
as alluded to above is, then, that the phase for a neutrino \MES, $j$, 
propagating non-radially through a Schwarzschild metric is purely 
proportional to its mass, $m_j$, squared 
[see Eq.~(58) of \cite{Fornengo1997} and 
also Eq.~(25) of \cite{Linet2002}].
This result is incorrect\footnote{Though it should be stressed here that
the authors of Ref. \cite{Fornengo1997} and many of the other papers we have mentioned
{\it do} obtain the correct result for the phase {\it difference} between neutrino \MES s 
traveling along the same macroscopic paths in curved spacetime (i.e., for
flat space neutrino oscillations in the presence of a point mass) because, in such cases,
any putative $\propto \ E$ term will vanish in subtracting one phase from the other (this term
 being the same for both phases).}: 
in the massless limit, the neutrino phase in curved spacetime should reduce 
(modulo spin-dependent corrections which vanish for radial trajectories 
\citep{Brill1957} and are negligible except in extreme, gravitational 
environments \citep{Alsing2001})
to the result for photons. 
And the photon phase is not zero 
(otherwise the interference fringes 
-- in space or energy -- 
predicted by the femto-lensing literature would not be produced), 
even though, of course, the classical action is zero along null geodesics. 
Indeed, the photon phase is essentially proportional to energy. As we show below,
furthermore, this $\propto \ E$ term is the leading order term for the neutrino phase
as well.

What has gone wrong when one's 
analysis misses the $\propto \ E$ term in the neutrino phase
 is that one has tried to simultaneously employ
two incompatible notions: 
the fundamentally wave or quantum mechanical idea of phase
with the particle notion of trajectory so that $x$ is given in terms of $t$ 
or {\it vice versa}. 
Even in the simpler case of flat space oscillations, the introduction 
(often implicitly)
of the idea of a trajectory 
-- or, more particularly, a group velocity --
into the calculation of neutrino phase leads to error 
(in particular, the recurring bugbear that the conventional formula 
for the neutrino oscillation length is wrong by a factor of two: 
see \cite{Giunti2002}). 
In the calculations set out below, we show the reader how the error 
of introducing a trajectory can be avoided. 
Furthermore, our method allows calculations to be performed along the actual 
(classical) 
paths
[not trajectories; i.e., we have $r(\phi)$, say, rather than $r(t), \phi(t)$] 
of the neutrino \MES s, rather than taking the approach of calculation
along the null geodesic employed in Ref \citep{Fornengo1997}.
On the other hand, our approach also circumvents the obligation to 
introduce extra phase shifts `by hand'. 
This artificial device becomes necessary when
one offsets either the emission times or positions of the different \MES s
with respect to each other so that 
they arrive at the same spacetime point 
(see \cite{Bhattacharya1999} for an example of this). 

The plan of this paper is the following: in \S\ref{section_beamsplitter}
we describe, for illustrative purposes, 
 interference of neutrino plane waves propagating through flat
space along both different (classical) paths and having, in general,
 different phase velocities. 
Then in \S\ref{section_curvednuphase} we
describe the calculation of the neutrino \MES \ phase in a 
Schwarzschild metric, correcting an erroneous result that has existed in 
the literature for some time. 
We then set out, in \S\ref{section_osnprob} 
the calculation of the analog of the
survival and oscillation probabilities in flat space
for neutrinos that have been
gravitationally lensed by a point mass. 
In \S(\ref{section_phenomenology}) we examine -- at an heuristic level --
questions of coherence that can effect the visibility of the 
GINI effect for neutrinos from supernovae. We give a proof-of-principle that 
the effect should be detectable.
In \S\ref{section_furtherwk}
we describe some limitations
of our method -- which stem particularly from the assumption of exclusively
classical paths -- and set out improvements to be made in further work. 
Finally, in an appendix, we 
set out a wave packet treatment of the neutrino beam splitter
toy model treated in \S 3 in terms of plane waves.
We derive results here pertaining to the analog of the
coherence length in conventional neutrino oscillations.

\begin{figure*}[ht]
\epsfig{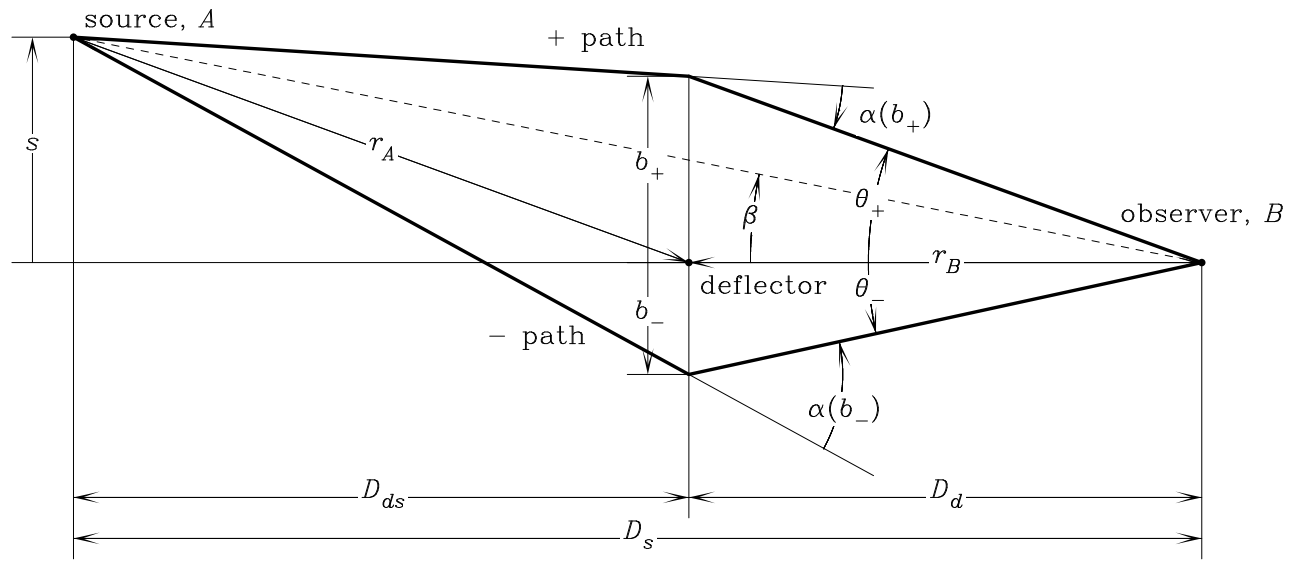}
\caption{The geometry of either a double slit interference experiment or a
two-image gravitational lens. The bold lines are the two neutrino paths,
$+$ and $-$, between the source, $A$, and the observer, $B$, and the
geometry is defined both in terms of physical variables ($r_A$, $r_B$,
$b_\pm$ and $s$) and astronomical/lensing variables ($D_d$, $D_s$,
$D_{ds}$, $\theta_\pm$ and $\beta$). Also included is the deflection
angle, $\alpha_{\pm}(b)$, parameterised as a function of the physical
impact parameter. It is implicit in this diagram that the deflector is
rotationally symmetric, and that the paths are thus confined to the plane
of the page defined by the source, deflector and observer.}
\label{figure:geometry}
\end{figure*}

\section[Neutrino Beam Splitter]
{Neutrino Beam Splitter}
\label{section_beamsplitter}

By way of an illustrative introduction to this topic we consider a toy model
of interference effects that can arise when there are both multiple paths
from a source to a detector, {\it {\` a} la} Young's double
slit experiment, and
different phase velocities
for the propagating particles, {\it {\` a} la} neutrino oscillations.
Of course, interference
requires that our experimental apparatus be unable to distinguish between
the  propagating particles (just as we likewise require for
interference that the apparatus is unable to identify {\it which} path
any single particle has propagated down). We therefore
require the propagating and detected particles to be different objects.
In this context, let us consider, for the sake of definiteness and relevance,
a {\it Gedanken Experiment} involving an imaginary 
(flat space) neutrino beam splitter in
the geometry illustrated in Fig.~\ref{figure:geometry}. We take it that 
the neutrinos' paths can be approximated as two
straight-line segments
along which momentum is constant in magnitude.
We need only treat, therefore,
one spatial dimension: $\int \mathbf{p}_j . \mathrm{d} \mathbf{x}
= \int {p}_j \mathrm{d} x 
= {p}_j \vert x \vert $, where $\vert x \vert$ is the total
distance along the two line segments.
We expect that the qualitative behavior of this device
shall illustrate many of the features expected to emerge
from interference of gravitationally lensed neutrinos.
Note that for reasons of clarity we present only a
plane wave treatment here, leaving a full wavepacket calculation for an
appendix. We stress, however, that wave packet considerations -- which 
allow, in particular, for a proper treatment of decoherence effects --
are, in general, important and must certainly be considered when
one is dealing with neutrinos that have propagated over long distances
(see \S 6). 

Let us write the ket associated with the 
neutrino flavour eigenstate $\alpha$ that has (in a loose sense)
propagated from
the source spacetime position $A = (x_A, t_A)$
to
detection position 
$B = (x_B, t_B)$ as
\begin{multline}
\label{eqn_bs}
\vert \nu_\alpha; A, B \rangle
= N \sum_p \sqrt{I_p} \sum_j U_{\alpha j} \\
\times \exp[-\mathrm{i} \Phi^p_j(E_j; L^{AB}_p, T^{AB})] 
\vert \nu_j \rangle,
\end{multline}
where $T^{AB}\equiv t_B - t_A$, $L^{AB}_p$ is the distance from 
source position $x_A$ to detector position $x_B$ along one of a finite number 
of paths labeled  by $p$, $\Phi^p_j(E_j; L^{AB}_p, T^{AB}) 
\equiv E_jT^{AB} - p_j(E_j)L^{AB}_p$, with
$p_j(E_j)$ denoting the momentum of \MES \ $j$ with energy $E_j$, and, finally,
$U$ is a unitary matrix relating the neutrino flavor eigenstates to the 
neutrino mass eigenstates. 
We have included the $\sqrt{I_p}$ factor to account for the fact that,
in general, we should allow for a path-dependence to the
amplitude. A situation where, for paths $p$ and $q$, $I_p \neq I_q$
 is the analog of a Young's slit experiment in which the slits are not equally
illuminated (thus reducing the {\it visibility} -- see
Eq.~(\ref{eqn_visibility}) -- of the resulting interference fringes).
The Schwarzschild lens scenario explored in \S\S 4 and 5 presents a situation
analogous to this: light or neutrinos propagating down the two classical paths
from source to observer will not be, in general, equally magnified.
Note that
we choose throughout this paper -- unless otherwise indicated -- to work in
units such that $\hbar = c = 1 \neq G$.  
The \MES s are assumed
to be on mass shell: $p_j(E_j) \equiv (E_j^2 - m_j^2)^{1/2}$.
The amplitude for a neutrino created as type $\alpha$ at the spacetime position
$A$ to be
detected as type $\beta$ at the spacetime position, $B$, 
of the detection event is then:
\begin{multline}
\label{eqn_bs'}
\langle \nu_\beta \vert \nu_\alpha; A, B \rangle 
= N \sum_p \sqrt{I_p} \sum_j U_{\alpha j} U_{\beta j}^* 
\\ \times \exp[-\mathrm{i} \Phi^p_j(E_j; L^{AB}_p, T^{AB})].
\end{multline}
Assuming a stationary source, we
can get rid of the unwanted dependence on time by averaging over $T^{AB}$
in the above to determine
a time-averaged oscillation probability {\it analog} 
at the detector position $x_B$ 
\cite{Beuthe2001,Beuthe2002,Giunti2003}.
This maneuver gives us that
\begin{multline}
\label{eqn_bs''}
\vert \langle \nu_\beta \vert \nu_\alpha; x_A,x_B \rangle \vert^2
\propto \int \mathrm{d}T 
\vert \langle \nu_\beta \vert \nu_\alpha; A,B \rangle \vert^2 \\
\propto \vert N \vert^2 \sum_{pq} \sqrt{I_pI_q} \sum_{jk} 
U_{\alpha j} U_{\beta j}^* U_{\beta k} U_{\alpha k}^*  \\
\times 
\exp[\mathrm{i}(p_j(E)L^{AB}_p - p_k(E)L^{AB}_q)],
\end{multline}
where we have $E_j = E_k \equiv E$ because
of the $\delta(E_j - E_k)$ that arises from the integration over time.

We find, then, after a simple calculation that the oscillation probability 
analog becomes
\begin{multline}
\label{eqn_bs'''}
\vert \langle \nu_\beta \vert \nu_\alpha; x_A,x_B \rangle \vert^2\\
\simeq \frac{1}{\sum_{rs} \sqrt{I_rI_s}} \sum_{pq} \sqrt{I_pI_q} \sum_{jk} 
U_{\alpha j} U_{\beta j}^* U_{\beta k} U_{\alpha k}^*  \\
\times  \exp\left[-\mathrm{i}\Delta \Phi_{jk}^{pq}\right],
\end{multline}
where the phase difference is given by
\begin{multline}
\label{eqn_splitterphase}
\Delta \Phi_{jk}^{pq} \equiv 
- E (L^{AB}_p - L^{AB}_q)
+ \left(\frac{m_j^2 L^{AB}_p - m_k^2 L^{AB}_q}{2 E}\right).
\end{multline}
Note that in Eq.~(\ref{eqn_bs'''})
the normalization has been determined by requiring
that 
$\vert \langle \nu_\beta \vert \nu_\alpha; x_A ,x_B \rangle \vert^2 \leq 1$.
The presence of the $\leq$ sign (as opposed to a simple equality)
comes about as the interference between states propagating down
different paths can result in minima at which the total neutrino 
detection probability is zero, in which case the usual unity 
normalisation is impossible (so now 
$\vert \langle \nu_\beta \vert \nu_\alpha; x_A,x_B \rangle \vert^2$
no longer has a direct interpretation 
as a
probability).
This is qualitatively different behavior
to that seen in neutrino oscillations where, given maximal mixing between
$\nu_\alpha$ and $\nu_\beta$,
an experiment
 might be conducted
in a position where only neutrinos of type
$\alpha$, say, are to be found or, alternatively, only of type
$\beta$,
but never in a position where none can be found in principle.

On the other hand the behavior explained above 
-- involving interference minima and maxima --
is obviously analogous to what one would expect in a double slit
experiment or similar. In fact, we shall show below that the phenomenology
of neutrino interference -- when there is more than one path from source
to detector -- is a convolution of the two types of interference outlined above.
This means that, in general, one cannot simply re-cast 
$\vert \langle \nu_\beta \vert \nu_\alpha; x_A ,x_B \rangle \vert^2$
in terms of a conditional probability, separating out the overall
interference pattern (with its nulls, etc.) from the conditional
probability that any {\it detected} neutrino has certain properties. To understand this
point, imagine setting all mixing angles to zero so that the \MES s and weak eigenstates 
are identical. The point now is that the interference patterns
for the various
detected weak/mass eigenstates are still different:
the phase difference (which now stems purely from
path difference) is dependent on the mass of the neutrino species involved. 
In other words, $\vert \langle \nu_\beta \vert \nu_\alpha; x_A ,x_B \rangle \vert^2$ 
will {\it not},
in general,
factorise into a conditional probability multiplied by an interference envelope because
the putative interference envelope is different for different \MES s. 
As we show below, however, in the ultra-relativistic limit, the cross term (that mixes
different path indices with different \MES \ indices) always turns out to be small
with respect to the other terms. In this limit, then, factorization is a good approximation.

We now consider two particular, illustrative cases of the neutrino 
beamsplitter {\it gedanken Experiment} for which we calculate relevant 
phase differences and oscillation probability
analogs.
\subsection{Two Path Neutrino Beam Splitter}
\label{section_planewavesplitter}

A particularly perspicuous 
example is given by the two-path example of the
above equations.
For this 
we specify 
a reference length, $L$, to which the two paths, of lengths $L_+$ and $L_-$, are related
by 
\be
L_\pm \equiv L \pm \frac{\Delta L}{2}.
\ee
This means that there are four phase difference types as labeled by the path indices
$p, q \in \{+,-\}$, viz:
\bea
i. \quad \Delta \Phi_{jk}^{++} &=& + \frac{\delta m^2_{jk} L + \Delta L/2}{2 E}\nn \\
ii. \quad \Delta \Phi_{jk}^{--} &=& + \frac{\delta m^2_{jk} L - \Delta L/2}{2 E} \nn \\
iii. \quad \Delta \Phi_{jk}^{+-} &=& -E \left(1 - \frac{m_j^2 + m_k^2}{4 E^2} \right)\Delta L
+ \frac{\delta m^2_{jk} L}{2 E} \nn \\
iv. \quad \Delta \Phi_{jk}^{-+} &=& +E \left(1 - \frac{m_j^2 + m_k^2}{4 E^2} \right)\Delta L
+ \frac{\delta m^2_{jk} L}{2 E}. \nn \\
\eea
In this case, then, Eq.~(\ref{eqn_bs'''}) becomes 
\begin{multline}
\label{eqn_planewave}
\vert \langle \nu_\beta \vert \nu_\alpha; x_A,x_B \rangle \vert^2\\
\shoveleft{
\simeq \frac{1}{I_+ + I_- + 2 \sqrt{I_+I_-}} 
}
\\
\times \sum_{jk} U_{\alpha j} U_{\beta j}^* U_{\beta k} U_{\alpha k}^* 
\exp\left(-\mathrm{i} \frac{\delta m^2_{jk} L}{2 E}\right) \\
\times \Biggl\{I_+ \exp\left(+\mathrm{i} \frac{\delta m^2_{jk} \Delta L}{4 E}\right) 
+ I_- \exp\left(-\mathrm{i} \frac{\delta m^2_{jk} \Delta L}{4 E}\right) \\
\shoveright{
+ 2 \sqrt{I_+I_-}\cos\left[E \left(1 -  \frac{m_j^2 + m_k^2}{4 E^2} \right)\right]\Biggr\}
\, .} \\
\end{multline}
This is an interesting result. It shows that the interference factorises into
a conventional, flat space oscillation term 
and an interference `envelope' in curly brackets
(as might be
expected from a Young's slit type experiment). If we now 
further particularize to the case
where $I_+ = I_-$, this envelope term becomes
\bea
\cos\left( \frac{\delta m^2_{jk} \Delta L}{4 E}\right) 
+ \cos\left[E \left(\!1 -  \frac{m_j^2 + m_k^2}{4 E^2} \right)\right] 
  \nn \\
= 
2 \cos \!
  \left[E \! \left(\!1 - \frac{m_j^2}{2 E^2} \right) 
  \! \frac{\Delta L}{2} \right] 
\cos \!
  \left[ E \! \left(\!1 - \frac{m_k^2}{2 E^2} \right) 
  \! \frac{\Delta L}{2} \right]
  ,\nn
\eea 
\be
\ee
which obviously reduces to 
the expected photon interference term $\propto \cos^2$ in the massless
limit. Later (in \S\ref{section_osnprob}) we shall see that this sort of
factorization property also arises, under a different set of assumptions, with
gravitational lensing of astrophysical neutrinos.

\subsection{Double slit experiment geometry}

Having considered
 a hypothetical neutrino plane wave beam-splitter 
(\S\ref{section_planewavesplitter}),
it is now possible to combine the rigorous, if simplistic, results
derived above
with heutristic arguments to investigate a more realistic
double slit neutrino interference experiment.
The choice of a double slit experiment is particularly relevant
not only because of its links with more familiar interference
phenomena, but also because a point-mass gravitational lens
admits two (significant) paths from source to observer.
Thus the results obtained in this section should provide a useful
guide to the qualitative behaviour of GINI in curved space time
considered in \S\ref{section_curvednuphase}.


The relevant geometry is illustrated in Fig.~\ref{figure:geometry}, 
showing the two
classical paths ($+$ and $-$) from source to detector, with the important
introduction of a more physically motivated set of lengths than the
reference length, $L$, used previously.
The entire experiment is
taken to be planar, with the coordinate system origin
on the line defined by the two slits.
(This is an arbitrary decision at present, but will coincide
with the position of the deflector in \S\ref{section_curvednuphase}.)
The slits are defined by their positions, $b_+$ and $b_-$
(which will correspond to image positions in 
\S\ref{sect_convl_lens_params});
the source position is given by both its radial coordinate, $r_A$,
and its perpendicular offset, $s$;
the observer position is defined by its radial coordinate, $r_B$,
alone.
There are several other plausible ways in which this geometry could
be defined, but all derived results become equivalent under the
assumption that $|b_\pm| \ll r_{\{A,B\}}$, as applied throughout.
Note also that the observer and source are interchangeable.

As discussed in \S\ref{section_beamsplitter}, 
all flat-space interference phenomena
can be treated in terms of path lengths.
The lengths of the two paths illustrated in 
Fig.~\ref{figure:geometry} 
are
\begin{eqnarray}
L_\pm 
  & = & \left( r_B^2 + b_\pm^2 \right)^{1/2}
        + \left[ r_A^2 - s^2 + (b_\pm - s)^2 \right]^{1/2} \\
  & \simeq & (r_A + r_B)
        \left[1 + \frac{1}{2 r_A r_B}
          \left( 
            b_\pm^2 - \frac{2 r_B}{r_A + r_B} 
            s b_\pm 
          \right)
        \right], \nonumber
\end{eqnarray}
where, in most cases, $b_-$ and $b_+$ are of opposite sign,
and the second line explicitly utilises
the fact that $|b_\pm| \ll r_{\{A,B\}}$.
The path difference is thus
\begin{eqnarray}
\Delta L_{+-} 
& = & L_+ - L_- \\
& \simeq & \frac{r_A + r_B}{2 r_A r_B}
    \left[
       b_+^2 - b_-^2 - \frac{2 r_B}{r_A + r_B}
       s \left(b_+ - b_-\right)
    \right] \nonumber
\end{eqnarray}
and the average path length is
\begin{eqnarray}
L & \simeq & (r_A + r_B) \\
& & 
        \left\{1 + \frac{1}{4 r_A r_B}
          \left[
            b_+^2 + b_-^2 
            - \frac{2 r_B}{r_A + r_B} s (b_+ + b_-)
          \right]
        \right\}. \nonumber
\end{eqnarray}

\subsection{Schwarzschild slit geometry}

In a laboratory-based double slit experiment
the two slit positions can be chosen arbitrarily,
but in the case of gravitational lensing
the impact parameters of the beams
are determined by a combination deflector and source
parameters (\S\ref{section_curvednuphase}).
Given the Schwarzschild metric around a point-mass 
(\S\ref{sect_convl_lens_params}),
the assumption of small deflection angles implies that
the source position and impact parameters are related by
[cf.\ Eq.~(\ref{equation:deltathetasq})]
\begin{equation}
\label{equation:b_sch}
b_+^2 - b_-^2 
= \Delta b_{+-}^2
  \simeq \frac{r_B}{r_A + r_B} s 
    \left(b_+ - b_-\right).
\end{equation}

Applying this result in the more general context of the double slit
experiment,
Eq.~(\ref{equation:b_sch}) can be rewritten as
\begin{equation}
\Delta L_{+-}
  = L_+ - L_- 
  \simeq - \frac{r_A + r_B}{2 r_A r_B}
    \left(b_+^2 - b_-^2\right).
\end{equation}
Similarly, the mass-length expression that appears in, e.g., 
Eq.~(\ref{eqn_splitterphase})
can be simplified to
\begin{eqnarray}
m_j^2 L_+ - m_k^2 L_-
& \simeq &
(r_A + r_B) \\
& & \mbox{} \times
\left[
\delta m^2_{jk} - \frac{1}{2 r_A r_B} (m_j^2 b_+^2 - m_k^2 b_-^2)
\right]. \nonumber
\end{eqnarray}

Substituting these expressions in 
Eq.~(\ref{eqn_splitterphase}) then gives the phase
difference between mass eigenstates $j$ and $k$ travelling down
paths $+$ and $-$ as
\bmu
\label{eqn_splitterphasepythag}
\Delta \Phi_{jk}^{+-} \simeq
+ E
 (r_A + r_B)\frac{\Delta b_{+-}^2}{2 r_A r_B} 
+\frac{\delta m^2_{jk}}{2 E} (r_A + r_B)\\
- \frac{r_A + r_B}{2 E}\frac{(m_j^2 b_+^2 - m_k^2 b_-^2)}{2 r_A r_B}.
\end{multline}
This expression includes contributions from both different phase
velocities and different path lengths, and can be understood
further by considering the special cases in which
(1) different mass neutrinos travel down the same path
or
(2) the same mass eigenstate travels down different paths.

\begin{enumerate}

\item From Eq.~(\ref{eqn_splitterphase}) for the general case
of the phase difference between the {\it same} \MES \
propagating along {\it different} paths we find
\bea
\label{eqn_splitterphase''}
\Delta \Phi_{jj}^{pq} &\equiv& 
- \left[E - \frac{m_j^2}{2E}\right](L_p - L_q) \nn \\
\eea
So that if we further particularize, as above, to paths passing through
slits at $b_+$ and $b_-$ we find
\bea
\label{eqn_samemes}
\Delta \Phi_{jj}^{+-}
& \simeq &  \left[E - \frac{m_j^2}{2E}\right]
\frac{(r_a + r_b)\Delta b_{+-}^2}{2 r_a r_b}.\nn \\
\eea
Notice in the above the similarity to the phase difference
for an ordinary
Young's double slit type experiment using photons,
namely,
\bea
\Delta \Phi^{pq}_\gamma &=&
-\bar{E}v(L_p - L_q) \nn \\
&=& - \left(\bar{E} - \frac{m^2}{2\bar{E}}\right)(L_p - L_q),
\eea
where $v$ is the (phase) velocity of the interfering particle
(which we assume to be relativistic).
We shall see below that the analog of this phase -- essentially
proportional to energy $\times$ path difference -- has been missed
in the existing literature on neutrino oscillations in curved space.
This has led to an incomplete result
suggesting
that the phase difference vanishes
in the massless limit even when there is more than one path from source
to detector.
\item Again for the general case,
for the phase difference between {\it different} mass eigenstates
propagating along the {\it same} path (i.e., the analog of the usual
phase difference encountered in neutrino oscillation experiments), we find
from Eq.~(\ref{eqn_splitterphase})
\bea
\label{eqn_splitterphase'}
\Delta \Phi_{jk}^{xx} &\equiv&
+ \left(\frac{\delta m_{jk}^2 L_x}{2 E}\right) \nn \\
& \simeq & \frac{\delta m_{jk}^2}{2 E}(L_a + L_b)
\left(1 + \frac{b_x^2}{2 L_a L_b}\right),
\eea
where $x \in \{p,q\}$.

\end{enumerate}

It is worth keeping the above expressions in mind
when considering the results for the neutrino phase difference
in curved spacetime presented in \S\ref{sect_convl_lens_params}.
As will be seen, the results obtained in this more complex physical
situation are analogous to those derived above,
e.g., compare Eq.~(\ref{eqn_samemes}) with 
Eq.~(\ref{samemesnonradschwarzphasediff}) and
Eq.~(\ref{eqn_splitterphase'}) with 
Eq.~(\ref{samepathnonradschwarzphasediff}).

\section{The Phase of a Neutrino Mass Eigenstate in Curved Spacetime}
\label{section_curvednuphase}

A neutrino beam splitter is in the realm of fantasy -- except
for the interesting case of gravitational  lensing of neutrinos: a
gravitational field can bring to a focus diverging neutrino beams, 
and  therefore
provide for multiple (classical) paths from a source to a detector.
In the remainder of this paper 
we explore whether any interesting, quantum mechanical interference
effects can arise in this sort of situation.

We shall be concerned below, therefore, with deriving an expression 
for the neutrino oscillation phase
in curved spacetime, in particular a Schwarzschild metric (this 
providing the simplest case in which gravitational lensing is possible).
Here we shall follow the 
development laid out in \cite{Fornengo1997} and \cite{Cardall1997} but,
importantly,
we shall also employ the prescription set out in \S\ref{section_beamsplitter}
that allows
for 
the
removal of time from consideration in the oscillation `probability' 
by integration over $T \equiv t_B - t_A$ where $A = (\bm{r}_A, t_A)$ 
and $B = (\bm{r}_B, t_B)$ are the
emission and detection events respectively \citep{Beuthe2001} . 
Note that we are assuming the semi-classical limit in which gravity is not quantized
and its effects can be described completely by a non-flat metric, 
$g_{\mu \nu} \neq \eta_{\mu \nu}$. 

The procedure we follow is to start with the
generalization of the equation for a \MES 's phase \ in flat spacetime to 
curved spacetime first arrived at by Stodolsky\citep{Stodolsky1979}:  
\begin{equation} 
\label{curvedphase}
\Phi_k(B,A) \qquad = \qquad \int_A^B p_{\mu}^{(k)} \mathrm{d}x^{\mu},
\end{equation}
where
\begin{equation}
\label{eqn_canmtm}
p_{\mu}^{(k)} \qquad = \qquad m_k g_{\mu \nu} \frac{\mathrm{d}x^{\nu}}{ds},
\end{equation}
is the canonically conjugate momentum to the coordinate $x^{\mu}$. 
Actually, as 
pointed out by Alsing et al. in Ref.~\cite{Alsing2001},
Stodolsky's expression for the phase is missing, in general, small
correction terms that arise from quantum mechanical 
modifications to the classical action. These
vary according to
the spin of the particle under consideration. Completely
fortuitously, the would-be
correction terms are identically zero
in the case of spin half  
particles in a {\it static} metric 
(whereas for particles with, e.g., spin zero or one they are non-zero)
so the 
 Stodolsky expression happens to be exact for the Schwarzschild metric and many
other cases of interest. Note in passing that this restriction to a static
metric means that this technology cannot -- as it stands -- treat, e.g., 
particle phases in a cosmological context.

We now introduce the metric of the Schwarzschild spacetime.
This may be written in radial 
co-ordinates, $x^\mu=(t,r,\vartheta,\varphi)$, as
\begin{equation} 
\label{schwarzmetric}
ds^2 \ = \ B(r) dt^2 - B(r)^{-1} dr^2 - r^2 d\vartheta^2 - r^2 
\sin^2 \vartheta d\varphi^2,
\end{equation}
where
\begin{equation}\label{B}
B(r) \ \equiv \ \left( 1 - \frac{2GM}{r} \right),
\end{equation}
and $G$ is the Newtonian constant and $M$ is the mass of the source of 
the gravitational
field, i.e., the lensing mass. 
Given the isotropy of
the gravitational field the motion of the neutrino mass eigenstate 
will be confined
to a plane which we take to be the equatorial one, $\vartheta = \pi/2$ and 
$d\vartheta = 0$.

The relevant components of the canonical momentum, Eq.~(\ref{eqn_canmtm}),
are, then \citep{Fornengo1997}:
\be
p^{(k)}_t = m_k B(r) \frac{dt}{ds},
\ee
\be
p^{(k)}_r = - m_k B^{-1}(r) \frac{dr}{ds},
\ee
and
\be
p^{(k)}_\varphi = - m_k r^2 \frac{d\varphi}{ds}.
\ee
These are all inter-related through the mass-shell condition
\citep{Fornengo1997}:
\begin{eqnarray}
\label{shell}
m^2_k &=& g^{\mu \nu} p_\mu^{(k)} p_\nu^{(k)} \nn \\
&=&
\frac{1}{B(r)} (p_t^{(k)})^2 - B(r) (p_r^{(k)})^2 -
\frac{(p_\varphi^{(k)})^2}{r^2}. \nn \\
\end{eqnarray}

Given that the components of the metric are independent
of the coordinates $t$ and $\varphi$, the momenta
associated with these quantities, $p^{(k)}_t$ and $p^{(k)}_\varphi$ 
shall be conserved along the
classical geodesic traced out by $\nu_k$. 
We define these constants of motion
as $E_k \equiv p^{(k)}_t$ and $J_k \equiv -p^{(k)}_\varphi$. These two are,
respectively, the energy and angular momentum seen by an observer at $r \to
\infty$
for the $k$th \MES \ \citep{Fornengo1997}.
They are {\it not} identical with the energy and angular momentum that
would be measured for $\nu_k$ at some definite, finite position $r$.
In general, however,
one may relate these quantities using the transformation law
that relates a local reference frame 
$\{x^{\hat{\alpha}}\} = \{\hat{t},\hat{r},\hat{\vartheta},\hat{\varphi}\}$
to the frame $\{x^\mu\} = \{t,r,\vartheta,\varphi\}$ \citep{Misner1973}:
\be
x^{\hat{\alpha}} = L^{\hat{\alpha}}_{\ \mu}x^\mu, \qquad \qquad
g_{\mu \nu} = L^{\hat{\alpha}}_{\ \mu} L^{\hat{\beta}}_{\ \nu} 
\eta_{\hat{\alpha} \hat{\beta}},
\ee
where the $L^{\hat{\alpha}}_{\ \mu}$'s are the coefficients of the
transformation between the two bases:
\bmu
\label{eqn_transform}
L^{\hat{t}}_{\ t} = \sqrt{\vert g_{tt} \vert}, \qquad
L^{\hat{r}}_{\ r} = \sqrt{\vert g_{rr} \vert}, \\
L^{\hat{t}}_{\ \vartheta} = \sqrt{\vert g_{\vartheta \vartheta} \vert}, \qquad
L^{\hat{\varphi}}_{\ \varphi} = \sqrt{\vert g_{\varphi\varphi } \vert}.
\end{multline}
So we have, in particular, that the local energy is given by
\citep{Cardall1997}:
\be
\label{eqn_locenergy}
E_k^{(loc)} (r) = |g_{tt}|^{-1/2} E_k = B(r)^{-1/2} E_k.
\ee

\subsection{Calculating the Phase Difference}

Given the above definitions, we now have that:
\bea 
\label{schwarzphase}
&\Phi_j^p(B,A) \qquad & = \qquad \int_A^B p_{\mu}^{(j)} \mathrm{d}x^{\mu}_p \nn \\
&&= \int_A^B [E_j dt - p_j(r)\mathrm{d}r_p - J_j d\varphi_p],\nn \\
\eea
where we have implicitly defined $p_j(r) \equiv - p_r^{(j)}$.
Note that we have explicitly introduced the path index $p$ which
allows for the possibility of
multiple paths from source to detector. Again, however, the integration
over $t$ is independent of the path as the  endpoints of this integration
are defined by the emission event and detection events. In fact, as discussed
above, $E_j$ is conserved over classical paths,
so that if \MES \ $j$ is assumed to travel down such a path, we can calculate
the phase it accumulated after leaving the source to be
\bea
\label{schwarzphase2}
&\Phi_j^p(B,A) & =  
\int_{t_A}^{t_B} E_j dt 
- \int_{r_A}^{r_B} 
\left[p_j(r) + J_j \left(\frac{d\varphi}{dr}\right)_j^p\right] \mathrm{d}r_p
\nn \\
&&= E_j (t_B - t_A) \nn \\
&& \qquad - \int_{r_A}^{r_B} 
\left[p_j(r) + J_j \left(\frac{d\varphi}{dr}\right)_j^p\right] \mathrm{d}r_p.
\eea
Of course, the quantity that governs the oscillation
phenomenology is the phase difference $\Delta \Phi_{kj}^{pq}$ 
where, generically,
interference can be between different \MES s and/or different paths
(cf.\ discussion in \S\ref{section_beamsplitter}). 
As things stand
this quantity would be parameterized in terms of both $t$ and $r$:
\begin{multline}
\label{schwarzphasediff}
\Delta \Phi_{jk}^{pq}(r_B,t_B,r_A,t_A)   =  
(E_j - E_k) (t_B - t_A) \\
- \Biggl\{\int_{r_A}^{r_B} 
\left[p_j(r) + J_j \left(\frac{d\varphi}{dr}\right)_j^p\right] \mathrm{d}r_p \\
- \int_{r_A}^{r_B} 
\Biggl[p_k(r) + J_k \left(\frac{d\varphi}{dr}\right)_k^q\Biggr] \mathrm{d}r_q\Biggr\}.
\end{multline}
We
therefore
follow the  prescription set out in \cite{Beuthe2001} to rid ourselves of the
unwanted time parameter: 
we assume a stationary source and integrate the interference term,
$\exp\left[ - \mathrm{i} \Delta\Phi_{kj}^{pq} \right]$,
over the unknown emission time $t_A$ (or, equivalently, the
transmission time $T \equiv t_B - t_A$). This results  
 in a very useful
$\delta(E_j - E_k)$. 

Note here that though the
 energies of different mass eigenstates are different
\citep{Giunti2001} -- 
so that the $\delta\left( E_k - E_j \right)$
arising from the time integration would
seem to imply no interference -- in fact,
in a correct treatment, massive neutrinos
are described by wave packets, not plane waves as here.
This means that, though  the average energies of different mass 
eigenstate wave packets are, in general, different,
 each massive neutrino wave packet has an energy spread
and
the detection process {\it can} pick up the same energy component
for different massive neutrinos
(see Refs.~\cite{Beuthe2001,Giunti2002d}).
If the energy spread of the wave packets is small
there is a suppression factor
that, formally, can only be calculated only with a wave packet treatment 
(cf.\ \S\ref{section_beamsplitter}), but, 
can also be assessed at the heuristic level 
(cf.\ \S\ref{section_phenomenology}).

Let us see how all the above works in practice.

\subsection{Radial Propagation}
\label{Radial Propagation}

We consider first the simple case of radial propagation, 
in which case there is a single classical path from source to
detector.
Along this path, the angular momentum vanishes and we have:
\begin{eqnarray}
\label{radialschwarzphasediff}
\Delta \Phi_{jk}(r_B,t_B,r_A,t_A)   
= (E_j - E_k) (t_B - t_A) \nn \\
- \int_{r_A}^{r_B} 
[p_j(r) - p_k(r)]\mathrm{d}r
\end{eqnarray}
We can determine $p_j(r) - p_k(r)$ from the mass-shell relation, 
Eq.~(\ref{shell})
\citep{Fornengo1997}:
\begin{eqnarray}
p_k(r) = \pm \frac{1}{B(r)}\sqrt{E_k^2 - B(r) m_k^2},
\end{eqnarray}
where the $+$ sign refers to neutrinos propagating outwards from the
gravitational well and the $-$ sign to neutrinos propagating inwards.
We can further simplify this relation by employing the binomial
expansion which, as in the flat space case, holds for relativistic
particles:
\bea
\label{eqn_binomcurved}
&\sqrt{E_k^2 - B(r) m_k^2} & \simeq E_k - B(r)\frac{m_k^2}{2E_k},
\eea
where $E_0$ is the energy at infinity for a neutrino \MES \ in the massless
limit 
(see \cite{Fornengo1997} for a detailed account of
the region of applicability of Eq.~(\ref{eqn_binomcurved})). 
We therefore have that
\be
p_j(r) - p_k(r) \simeq \pm \frac{1}{B(r)}(E_j - E_k)  
\mp \left(\frac{m^2_j}{2E_j} - \frac{m^2_k}{2E_k}\right).
\ee
The phase difference then becomes
\begin{multline}
\label{radialschwarzphasediff2}
\Delta \Phi_{jk}(r_B,t_B,r_A,t_A) \\
\simeq (E_j - E_k) \left[(t_B - t_A) 
\mp \int_{r_A}^{r_B}\frac{\mathrm{d}r}{B(r)}\right]
\\
+ \left(\frac{m^2_j}{2E_j} - \frac{m^2_k}{2E_k}\right)\vert r_B - r_A \vert. 
\end{multline}
Given  the oscillation `probability' shall be, following our
previously-establsihed procedure (cf.\ \S\ref{section_beamsplitter}),
 integrated over $T \equiv t_B - t_A$,
the {\it relevant} phase difference
can be seen to be
\bea
\label{radialschwarzphasediff3}
\Delta \Phi_{jk}(r_B,r_A)
&\simeq&
\frac{\delta m_{jk}^2}{2E_\nu} \left| r_B - r_A \right| \nn \\
&\simeq&\frac{\delta m_{jk}^2}{2E_0} \left| r_B - r_A \right|
\,.
\eea
where $E_\nu = E_j = E_k$ and
$E_0$ is the energy at infinity for a massless particle and, as in flat
space, the following relation holds \citep{Fornengo1997}:
\be
E_k \simeq E_0 + {\cal O}\left(\frac{m_k^2}{2E_0}\right).
\ee

To digress a little, note that 
the result presented in Eq.~(\ref{radialschwarzphasediff3}), 
arrived at previously 
\citep{Bhattacharya1996,Fornengo1997,Cardall1997,Bhattacharya1999},
must be interpreted with some care:
in
Eq.~(\ref{radialschwarzphasediff3})
one must keep in mind that the radial distance $\left| r_B - r_A \right|$ is
a {\it coordinate distance}, and not the proper distance the various
\MES s experience (except in the flat space case to which Eq.
(\ref{radialschwarzphasediff3}) clearly reduces in the limit
of a vanishing lensing mass) and that
$E_0$ does not represent a locally-detected energy.
Following \cite{Fornengo1997}, however, we can convert the phase difference
so that it appears in terms of these parameters.
The proper distance is given by (cf.\ Eq.~(\ref{eqn_transform})):
\bea
\label{eqn_propdist}
&L_\mathrm{prop} &\equiv \int^{r_B}_{r_A} \sqrt{g_{rr}} \mathrm{d}r \\
&& \simeq r_B - r_A + GM\ln\frac{r_B}{r_A},
\eea 
where in the second line we have assumed the weak field limit holds.
This demonstrates that, in a gravitational field, 
the length relevant to the calculation of phases, 
$\vert r_B - r_A \vert$, is actually shorter than the
distances experienced by the propagating particles, $L_\mathrm{prop}$.
Substituting Eqs.~(\ref{eqn_locenergy}) and (\ref{eqn_propdist}) into
(\ref{radialschwarzphasediff3}) we determine that (cf.\ \citep{Fornengo1997}):
\begin{eqnarray}
\label{phi_measured}
\Delta\Phi_{jk}(r_B,r_A) 
\hspace{-3pt}
&\simeq& 
\hspace{-3pt}
\left(
\frac{ \Delta m^2_{jk} L_\mathrm{prop}}{2 E_0^{(loc)}(r_B)}
\right) 
\nn \\
&&\times \left[
1
-
G M
\left(
\frac{1}{L_\mathrm{prop} }
\,
\ln \frac{r_B}{r_A}
-
\frac{1}{r_B}
\right)
\right]
\;. \nn \\
\end{eqnarray}

\subsection{Non-Radial Propagation}
\label{Nonradial Propagation}

We turn now to the more interesting case presented by non-radial propagation.
Here there will be, generically, more than one path for the \MES s to take
from source to detector and we have the possibility, therefore, of 
interference between particles on these different paths.

The phase difference we must calculate is given by
Eq.~(\ref{schwarzphasediff}). To proceed with this calculation
we must determine a value
for 
\begin{equation*}
p_j(r) + J_j \left(\frac{d\varphi}{dr}\right)_j^p.
\end{equation*}
Firstly recall that $J_j$ is constant along the classical path taken by
$\nu_j$. 
Now, using the fact (see \cite{Landau1975}, Eq.(101.5)) that
\begin{eqnarray}
\label{eqnmassangmtm}
\left(\frac{\mathrm{d}\phi}{\mathrm{d}r}\right)_j^p
&=& 
\pm \frac{J_j^p}{r^2 \sqrt{E_j^2 
- \left(m_j^2 + \frac{(J_j^p)^2}{r^2}\right) B(r)}} \nonumber \\ 
&\simeq&
\pm \frac{b_p \left(1 - \frac{m_j^2}{2 E_j} \right)}
{r^2 \sqrt{1 
- \frac{B(r)}{E_j^2}\left[m_j^2 + 
\frac{E_j^2 b_p^2}{r^2}\left(1 - \frac{m_j^2}{E_j^2}\right) \right]}},
\nn \\
\end{eqnarray}
and, given that from the mass-shell relation we have 
[see Eq.~(49) of \cite{Fornengo1997}]:
\begin{eqnarray}
B(r)p_j(r) &\simeq& \pm \sqrt{1 - B(r)\frac{b_p^2}{r^2}} \nn \\
&&\times
\left[1 - \frac{B(r)(1 - b_p^2/r^2)}{1 - B(r)b_p^2/r^2}
\frac{m_j^2}{2 E^2_j}\right],
\end{eqnarray}
where $b_p$ is the impact parameter for path $p$, one may determine that:
\begin{multline}
\label{eqnsunnyday}
p_j(r) + J_j^p \left(\frac{\mathrm{d}\phi}{\mathrm{d}r}\right)_j^p 
\simeq
\pm \frac{E_j}{B(r)\sqrt{1 - B(r)\frac{b_p^2}{r^2}}} \\
\mp \frac{m_j^2}{2E_j^2} \frac{1 
+ (1 - 2B(r))\frac{b_p^2}{r^2}}{\left(1-B(r)\frac{b_p^2}{r^2}\right)
^{\frac{3}{2}}}. 
\end{multline}
In the above we have also employed the fact that the angular momentum
of \MES \ $j$ (traveling along path $p$)
at infinity is 
given in terms of $\nu_j$'s energy at infinity, $E_j$, the impact parameter
along the path being considered, $b_p$, and 
$\nu_j$'s velocity at infinity, $v_j^{(\infty)}$ \citep{Fornengo1997}:
\bea
&J_j^p &= E_j b_p v_j^{(\infty)} \nn\\
&& \simeq E_j b_p \left(1 - \frac{m_j^2}{2E_j^2}\right).
\eea
We can further evaluate Eq.~\ref{eqnsunnyday} by replacing
the path-dependent impact parameter, $b_p$, with the minimal radial co-ordinate
for the same path, $r_0^p$. The relation between these two
is found  by noting that at the position of closest approach the rate of change
of the co-ordinate $r$ with respect to the angle $\phi$ vanishes
\citep{Fornengo1997}. For the massive case (Eq.~\ref{eqnmassangmtm}), this 
implies that:
\begin{eqnarray}
\label{eqnbpro}
b_p^2 \simeq \frac{1 + \frac{m_j^2}{E_j^2} \frac{2 GM}{r^p_0}}{B(r^p_0)} \,.
\end{eqnarray}
Employing Eq.~(\ref{eqnbpro}), taking the weak field limit, and
also expanding to ${\cal O}(m_j^2/E_j^2)$ we find that:
\bmu
\label{eqncontdsun}
p_j(r) + J_j^p \left(\frac{\mathrm{d}\phi}{\mathrm{d}r}\right)_j^p \\
\simeq \pm E_j \left[ \frac{r}{\sqrt{r^2 - (r^p_0)^2}} 
+ GM \frac{2r - 3r^p_0}{\sqrt{r^2 - (r^p_0)^2}(r + r^p_0)} \right] \\
 \mp \frac{m_j^2}{E_j}  \left[ \frac{r}{\sqrt{r^2 - (r^p_0)^2}} 
-GM \frac{r^p_0}{\sqrt{r^2 - (r^p_0)^2}(r + r^p_0)} \right] \\
\end{multline}
With this result in hand, 
we can complete the calculation of Eq.~(\ref{schwarzphase2}),
the phase accumulated by \MES \ in non-radial propagation from spacetime
position $A = (r_A, t_A)$ to $B = (r_B, t_B)$, where either
$r_A$ or $r_B$ is the minimal radial co-ordinate encountered over the
journey (i.e. the path is either non-radially inwards {\it or} outwards
but not both). After an elementary integration we find that
\begin{multline}
\Phi_j^p(B,A)  
\simeq E_j (t_B - t_A)\\
\mp E_j \left[
\sqrt{r_B^2 - (r^p_0)^2} - \sqrt{r_A^2 - (r^p_0)^2}  \right. \\
 + 2 GM \ln \left(\frac{r_B + 
\sqrt{r_B^2 -(r^p_0)^2}}{r_A + \sqrt{r_A^2 -(r^p_0)^2}} \right)  \\
  \left.  
\shoveright{+GM \left(\sqrt{\frac{r_B - r^p_0}{r_B + r^p_0}} - 
\sqrt{\frac{r_A - r^p_0}{r_A + r^p_0}} \right) \right]}  \\
\pm \frac{m_j^2}{2 E_j} \left[\sqrt{r_B^2 - (r^p_0)^2} 
- \sqrt{r_A^2 - (r^p_0)^2}  \right. \\
 \left.\qquad \qquad -GM \left(\sqrt{\frac{r_B - r^p_0}{r_B + r^p_0}} - 
\sqrt{\frac{r_A - r^p_0}{r_A + r^p_0}} \right) \right],  \\
\label{schwarzphaseblah}
\end{multline}
where  the upper signs pertain if $dr$ is positive (outward propagation)
and the lower
if $dr$ is negative.

\subsection{Neutrino Lensing}
\label{section_nulensing}

Finally let us consider the case of gravitational lensing of neutrinos. In
this case the neutrinos propagate non-radially 
along classical paths, labelled by index $p$, from radial position
$r_A$, inwards to a path-dependent minimal radial co-ordinate $r_0^p$,
and outwards again to a detector situated at radial co-ordinate
$r_B$. As presaged above,
in this situation there will be (at least potentially)
interference not only between
different \MES s propagating down the same classical path, but also between
\MES s propagating down different paths ($p$ and $q$, say).
Taking into account the sign of the momentum along these two legs, 
we find, following the developments above, 
that the relevant phase is given by
\begin{multline}
\label{nonradschwarzphase}
\Phi_j^{p}(B,A) \\
\shoveleft{
\simeq E_j (t_B - t_A)
} \\
- E_j \left[ \sqrt{r_A^2 - (r^p_0)^2} +
2 GM \ln \left(\frac{r_A + 
\sqrt{r_A^2 -(r^p_0)^2}}{r^p_0} \right)  \right.\\
  \left. 
\shoveright{+ GM \left(\sqrt{\frac{r_A - r^p_0}{r_A + r^p_0}}\right) 
+ (r_A \to r_B)\right]}
 \\
\shoveleft{+ \frac{m_j^2}{2 E_j} \Biggl[\sqrt{r_A^2 - (r^p_0)^2} 
}\\
\shoveright{- GM \left(\sqrt{\frac{r_A - r^p_0}{r_A + r^p_0}}\right) + (r_A \to r_B) 
\Biggr],}\\
\end{multline}
where $(r_A \to r_B)$ mean add another term of the same form but with $r_A$ replaced with $r_B$.

Before proceeding any further with the calculation it behooves us here to
establish the plausibility of Eq.~(\ref{nonradschwarzphase}) by 
showing its relation to results known from some simpler cases.
In the $M \to 0$ limit this equation becomes
\bea
\label{eqnMzero}
\Phi_j^{p}(B,A)&\simeq&
E_j (t_B - t_A)
- E_j\left(1 - \frac{m_j^2}{2 E_j^2}\right) \nn \\
&& \qquad \times
\left[ \sqrt{r_A^2 - (r^p_0)^2} + \sqrt{r_B^2 - (r^p_0)^2}\right]. \nn \\ 
\eea                                                           

Now refer back to Fig.~\ref{figure:geometry},
and take the coordinate origin on the diagram to denote
the position of a lensing point mass. 
In the massless case, the two classical paths reduce to the single
`undeflected' path denoted by the dashed line in the diagram.
Denote the minimal radial coordinate along this path by $r_0$ (which intersects
the dashed line at right angles).
Clearly, then, the geometrical 
length of the path from source to detector is
$\sqrt{r_A^2 - r_0^2} + \sqrt{r_B^2 - r_0^2}$. Now, given we 
know that in flat space the 
phase of \MES \ $k$ is given by Eq.~(\ref{curvedphase}) 
with Minkowski metric, viz: 
\begin{equation} 
\label{flatphase}
\Phi_j \qquad = \qquad E_j(t_B - t_A) - \bm{p}_j.(\bm{x}_B - \bm{x}_A).
\end{equation}
then,
for the  $M \to 0$ case
illustrated in Fig.~\ref{figure:geometry}
this becomes
\begin{eqnarray}
\label{flatphase2}
\Phi_j  
&\simeq&  E_j(t_B - t_A) - E_j\left(1 - \frac{m_j^2}{2 E_j^2}\right)\nn \\
&& \qquad \qquad \times
\left( \sqrt{r_A^2 - r_0^2} + \sqrt{r_B^2 - r_0^2}\right). \nn \\
\end{eqnarray}
With Eq.~(\ref{flatphase2})
we have, then independently established the plausibility
of Eq.~(\ref{eqnMzero}), once one takes into account the fact that,
in the massless lens case, all classical paths
converge on the same undeflected path (as mentioned above) so that
in this limit $ r_0^p =  r_0^q \equiv r_0$ $ \forall p,q$.

The other limit of interest is to take
$m_j \to 0$ in Eq.~(\ref{nonradschwarzphase}).
In doing this -- and then setting the temporal and spatial
contributions to the phase equal as appropriate for
a null geodesic -- we find that we have re-derived the
Shapiro time delay 
[see, e.g., Eq.~(8.7.4) of Ref.~\citep{Weinberg1972}].

Continuing with our main calculation, we can re-write 
Eq.~(\ref{nonradschwarzphase})
in terms of $b_p$ by inverting Eq.~(\ref{eqnbpro}). 
If we also expand to
${\cal O}(b_p^2/r^2_{B \leftrightarrow A})$, we find that 
\begin{multline}
\label{nonradschwarzphase3}
\Phi_j^{p}(B,A)  \\
\shoveleft{\simeq
E_j (t_B - t_A) 
}\\
\shoveleft{
\qquad - E_j (r_A + r_B)
} \\
 \times \left\{ 1 - \frac{b_p^2}{2 r_A r_B} + \frac{2 GM}{r _A + r_B}
\left[ 1 + \ln\left(\frac{4 r_A r_B}{b_p^2} \right)\right] \right\} \\
\shoveright{
+ \frac{m_j^2}{2 E_j}(r_A + r_B) \left(1 - \frac{b_p^2}{2 r_A r_B} 
-  \frac{2 GM}{r _A + r_B} \right).
}\\
\end{multline}
We can now find the phase difference, which allows for interference between
different paths and/or different \MES s, by the  usual
integration over $T$ (so that we have $E_\nu = E_j = E_k$):
\begin{multline}
\label{nonradschwarzphasediff}
\Delta\Phi_{jk}^{pq}(r_B,r_A) \\
\simeq
+ E_\nu (r_A + r_B)\left(\frac{\Delta b_{pq}^2}{2 r_A r_B} 
+ \frac{4 GM}{r _A + r_B} \ln\left|\frac{b_p}{b_q}\right| \right)  \\
+ \frac{\delta m_{jk}^2}{2 E_\nu}(r_A + r_B)
 \left(1 - \frac{2 GM}{r_A + r_B}\right) \\
- \frac{r_A + r_B}{2 E_\nu} 
\left(\frac{m_j^2b_p^2 - m_k^2b_q^2}{2r_Ar_B}\right),
\\
\end{multline}
where $\Delta b^2_{pq} \equiv b^2_p -b^2_q$ and, in our notation,
 $\Delta\Phi_{jk}^{pq}(r_B)$ denotes the phase difference between
\MES \ $j$ traveling down path $p$ and \MES \ $k$ traveling down path $q$
\footnote{Note that the $M \to 0$ case of this equation
can be re-derived by, again, considering Fig.~\ref{figure:geometry}
while noting, in particular, that the Schwarzschild lens satisfies
Eq.~\ref{equation:b_sch}
for the two classical paths.}. 
Eq.~(\ref{nonradschwarzphasediff}) is one of the major results of this paper.
Note that the presence of the $\propto E_\nu$ term in this equation 
-- missed in Ref.~\citep{Fornengo1997} -- 
ensures that the phase difference behaves properly
in the massless limit (i.e., does {\it not} vanish).
In passing, also
note that the above equation satisfies
the discrete 
 symmetry of swapping $B$ and $A$, as it should: the same result must be 
obtained for the phase difference (in a static spacetime) if we swap the
positions of source and observer.

Also recall that, excluding the case of perfect alignment of source, lens,
and observer, there are only two possible classical paths from source to
observer for the
Schwarzschild case. These we label by $+$ (this path having an impact 
parameter somewhat greater than the impact parameter
for an undeflected ray) and $-$ (this path  having an impact
parameter on the `opposite' side of the lens to the undeflected ray).
We require, therefore, that $p,q \in \{+,-\}$ and
in the particular case 
that we are considering interference between the same
mass eigenstates propagating down different paths 
Eq.~(\ref{nonradschwarzphasediff}) becomes
\begin{multline}
\label{samemesnonradschwarzphasediff} 
\Delta\Phi_{jj}^{+-}(r_B,r_A)\\
\simeq
+ E_\nu (r_A + r_B)\left(\frac{\Delta b_{+-}^2}{2 r_A r_B} 
+ \frac{4 GM}{r _A + r_B} \ln\left|\frac{b_+}{b_-}\right| \right) \\
- \frac{m_j^2}{2 E_\nu} (r_A + r_B)
\frac{\Delta b_{+-}^2}{2r_Ar_B}.\\
\end{multline}
Alternatively, in the case of 
different \MES s traveling down the same path 
(i.e., `ordinary' neutrino oscillations, but in curved space),  
Eq.~(\ref{nonradschwarzphasediff}) 
becomes
\begin{multline}
\label{samepathnonradschwarzphasediff}
\Delta\Phi_{jk}^{pp}(r_B,r_A)  
\simeq 
\frac{\delta m_{jk}^2}{2 E_\nu}(r_A + r_B) \\
 \times 
\left(1 - \frac{b_p^2}{2 r_A r_B} - \frac{2 GM}{r_A + r_B}\right).
\end{multline}
Note that Eqs.~(\ref{nonradschwarzphasediff}), 
(\ref{samemesnonradschwarzphasediff}), and 
(\ref{samepathnonradschwarzphasediff}) give us that
\begin{multline}
\Delta\Phi_{jk}^{pq}(r_B,r_A) = \Delta\Phi_{jj}^{pq}(r_B,r_A) +
\Delta\Phi_{jk}^{pp}(r_B,r_A) \\  
+ {\cal O}\left[\frac{\delta m_{jk}^2}{2 E_\nu} (r_A + r_B) 
\frac{\Delta b_{pq}^2}{2 r_A r_B}\right].
\end{multline}
This correction term will be small with respect to other terms (given our assumptions
of ultra-relativistic neutrinos and undeflected impact parameters small with 
respect to the overall distances between source-lens and lens-observer). 
In fact, the third term of Eq.~(\ref{nonradschwarzphasediff}) can be expected to be
suppressed with respect to the first term by ${\cal O}(m^2/E^2_\nu)$ and with
respect to the second term by ${\cal O}(b^2/(r_A r_B))$.
The consequence of this is that the phase may be written
\be
\label{eqnseparable}
\Delta \Phi^{pq}_{jk} 
\simeq \Delta \Phi_{jk} +  \Delta \Phi^{pq},
\ee
satisfying what we label `separability',
where
\bea
\label{eqn_phasediff_pq}
\Delta \Phi^{pq} \equiv 
E_\nu (r_A + r_B)\left(\frac{\Delta b_{pq}^2}{2 r_A r_B} 
+ \frac{4 GM}{r _A + r_B} \ln\left|\frac{b_p}{b_q}\right| \right) \nn \\
- \frac{\overline{m}^2}{2 E_\nu} (r_A + r_B)
\frac{\Delta b_{pq}^2}{2r_Ar_B}.
\nn \\
\eea
and
\bea
\label{eqn_phasediff_jk}
\Delta \Phi_{jk} &\equiv& 
\frac{\delta m_{jk}^2}{2 E_\nu}(r_A + r_B)
 \left(1 - \frac{\overline{b}^2}{2 r_A r_B} - \frac{2 GM}{r_A + r_B}\right),
\nn \\
\eea
where 
\bea
\overline{m} \equiv \frac{1}{N_\nu}\sum^{N_\nu}_j m_j
&\textrm{and} & \overline{b} \equiv \frac{1}{N_\textrm{path}}\sum^{N_\textrm{path}}_j b_p,
\eea
with $N_\nu$ the number of neutrino \MES s and $N_\textrm{path}$ the number of
classical paths from source to detector (two in the case of the Schwarzschild metric).
What Eq.~(\ref{eqnseparable}) says in words is that the phase difference that
develops between source and detector is due to 
two effects that can be considered separately: (i)
a phase difference -- independent  of which \MES \ is under consideration -- 
that develops because
of the different lengths of the paths involved and (ii) the phase difference that develops
because the different \MES s travel with different phase velocites. 
This situation is analogous to two runners who run along two very similar -- though not
identical -- paths, with similar -- though not
identical -- velocities: to first order, the difference in the finishing times between the two
depends on terms proportional to the difference in lengths of the two courses, $\Delta L$,
 and the difference
in the runners' velocities, $\Delta v$, but not, by definition, on terms 
$\propto  \Delta L \, \Delta v$.


\subsection{Phase Difference in Terms of Conventional Lensing Parameters}
\label{sect_convl_lens_params}

To facilitate interpretation of the above results 
in an astrophysical context
-- and, eventually, to introduce an evolving cosmological model --
it is useful to re-express the phase difference in the language
of standard gravitational lensing theory
(despite the fact that the particles being lensed are not photons).

\subsubsection{The lens equation}

The classically allowed neutrino paths in the presence
of a deflector can be derived by 
reconsidering the geometry shown in Fig.~\ref{figure:geometry}.
Under the assumption that $|b_\pm| \ll r_{\{A,B\}}$,
the source offset, $s$, can be related to
the impact parameter, $b$, by the lens equation:
\be
\label{eqn_lensgeneral}
s \simeq \frac{r_B}{r_A + r_B} b + r_A \, \alpha(b),
\ee
where $\alpha(b)$ is the deflection angle of the lens as a function of
impact parameter.

It is standard practice to reexpress the lens equation in terms 
of angular variables. 
This entails replacing the source offset and impact parameters
with angles (on the sky of the observer)
and radial coordinates with line-of-sight distances.
These conversions are summarised graphically in Fig.~\ref{figure:geometry},
which leads to the following replacements:
$r_A \to ({D_{ds}^2 + s^2})^{1/2} \simeq D_{ds}$,
where $D_{ds}$ is the distance from deflector to source;
$r_B \to D_d$, where $D_d$ is the distance from observer to deflector;
and $r_A + r_B \to D_s$,
where $D_s$ is the distance from observer to source.
The notation employed for the distance measures is suggestive of their
being the angular diameter distances used to relate angles and lengths
in an evolving cosmological model, and they fulfil an analogous role here.
It is most important to note, however, that they are
not true angular diameter distances and
the following results are only quantitatively valid on scales sufficiently
small that the expansion of the Universe can be ignored
(e.g., the Milky Way or the Local Group).
These results will be extended to an evolving cosmology in \citep{Crocker2003}.

The above caveats notwithstanding, the angular position
(relative to the deflector) of an image
with impact parameter $b$ is now simply
\be
\theta \simeq \frac{b}{r_B} \simeq \frac{b}{D_d},
\ee
and the position of the source can be given in terms of
an unobservable angular parameter $\beta$ as
\be
\beta \simeq \frac{s}{r_A + r_B} \simeq \frac{s}{D_s}.
\ee
Inserting these definitions into Eq.~(\ref{eqn_lensgeneral}),
the lens equation becomes
\begin{equation}
\label{equation:lens_angle}
\beta \simeq \theta + \frac{D_{ds}}{D_s} \alpha(D_d \theta).
\end{equation}
The position(s) of the images formed by a source in a 
given position can then be found 
for a given choice of deflector model.

\subsubsection{The Schwarzschild lens}

In a Schwarzshild metric, the total angular
deflection of a particle of mass $m$ impinging on a point-mass $M$ 
with undeflected impact parameter $b$ is 
(see, e.g., \cite{Schneider1992}):
\be
\alpha_{\rm{gen}}(b) = - \frac{4 GM}{b} \frac{1}{2}
  \left(1 + \frac{1}{v_\infty^2} \right),
\ee
where
$v_\infty$ is the particle's speed at an infinite distance from the mass
and it has been assumed that $b \gg 2 G M = R_S$, the deflector's
Schwarzschild radius.
For an ultra-relativistic particle, this becomes
\be
\label{eqn_defln}
\alpha_{\rm{rel}}(b) \simeq 
  - \frac{4 GM}{b}\left(1 + \frac{m^2}{2 E^2} \right),
\ee
where $E$ is its coordinate energy 
(equal to the energy measured at an infinte distance 
from the mass). 

For astrophysical neutrinos, however, $m^2 / (2 E^2) \ll 1$
and so it is an excellent approximation to assume they travel
along classical photon paths, for which 
\be
\label{eqn_angdefln}
\alpha_{\rm{light}}(b) = - \frac{4 GM}{b}.
\ee
Previously we have been rigorous in taking the classical 
paths of massive particles 
from source to observer but, as will be seen below, 
this assumption is entirely self-consistent when dealing with
weak-field gravitational effects and ultra-relativistic particles.
Note also that a corollary of this approximation is that the different 
mass eigenstates are assumed to travel down identical paths 
(whereas in reality the heavier eigenstates will fall marginally
deeper into the deflector's potential well).

Applying the above deflection law to
Eq.~(\ref{equation:lens_angle}) gives the point-mass lens equation as
\be
\label{equation:beta_theta}
\beta \simeq \theta - \frac{\theta_E^2}{\theta}
\ee
where 
\be
\label{equation:theta_e}
\theta_E = \sqrt{4 G M \frac{D_{ds}}{D_d D_s}}
\ee
is the Einstein radius of the lens.
This is the angular radius of the circular image that would be formed
in the case of perfect source-deflector-observer alignment
(i.e., $\beta = 0$) and thus depends on distance factors
as well as the lens mass.
Solving the lens equation then gives the image positions as
\be
\theta_\pm
  \simeq 
  \frac{1}{2} \left( \beta \pm \sqrt{\beta^2 + 4 \theta_{\rm{E}}^2} \right).
\ee
This also implies the useful Schwarzschild-specific result that 
\be
\label{equation:deltathetasq}
\theta_+^2 - \theta_-^2 = 
\Delta \theta^2_{+-}
  \simeq \beta (\theta_+ - \theta_-).
\ee

Having found a relationship between the angular position of a neutrino
source and its images, the expression for the phase
difference given in Eq.~(\ref{nonradschwarzphasediff}) 
can be recast in a form containing 
only line-of-sight distances and angular variables.
This yields the Schwarzschild-specific result that 
\begin{eqnarray}
\label{nonradschwarzphasediffdimless2}
\Delta\Phi_{jk}^{pq} & \simeq &
E_\nu \,
\frac{D_d D_s}{D_{ds}} \left[\frac{\Delta \theta^2_{pq}}{2} 
+ \theta_{\rm{E}}^2 \,\ln 
  \left(\left|\frac{\theta_p}{\theta_q}\right|\right) 
\right] \nonumber \\
& + & \frac{\delta m_{jk}^2}{2 E_\nu} D_s 
  \left(1 - \frac{D_d}{D_{ds}} \theta_{\rm{E}}^2 \right) \nonumber \\
& - & 
\frac{1}{2 E_\nu} \frac{D_d D_s}{D_{ds}}
\frac{m_j^2 \theta_p^2 - m_k^2 \theta_q^2}{2} .
\end{eqnarray}
The second term in this equation is simply the phase difference that
develops between mass eigenstates $j$ and $k$ traveling along 
the same path for distance $D_s$, 
with a small correction for the presence of the deflector.
The first term encodes the path difference along the trajectories
$p$ and $q$, with separate contributions from the geometrical effect
($\propto \Delta \theta_{pq}^2$) 
and the reduced coordinate velocity close to the deflector
[$\propto \ln(|\theta_p / \theta_q|)$].
The final cross term is the leading order contribution 
from different eigenstates traveling down different paths.
From the discussion in the previous section, 
this term will be small in general.

Given that interference effects can only ever be important 
when the detector cannot resolve different image positions
(i.e., it cannot know down which path the neutrino has travelled),
having the phase difference in terms of $\theta_p$ and $\theta_q$
is not as useful as expressing it as a function of the (angular) source
position, $\beta$.

For the Schwarzschild lens the conversion from $\theta$ to $\beta$
is given in Eq.~(\ref{equation:beta_theta}), and substituting this into 
Eq.~(\ref{nonradschwarzphasediffdimless2}) then gives 
(for \MES \ $j$ down path $+$ and \MES \ $k$ down path $-$)
\begin{eqnarray}
\label{nonradschwarzphasediffdimless'}
\Delta\Phi_{jk}^{+-} 
& \simeq &
E_\nu \frac{D_d D_s}{D_{ds}} \Biggl[
    \frac{\beta \sqrt{\beta^2 + 4 \theta_{\rm{E}}^2}}{2} \nonumber \\
& &    \qquad  \qquad \qquad + \theta_{\rm{E}}^2 \ln \left( \left|
      \frac{\beta + \sqrt{\beta^2 + 4 \theta_{\rm{E}}^2}}
           {\beta - \sqrt{\beta^2 + 4 \theta_{\rm{E}}^2}} 
      \right| \right) \Biggr] \nonumber \\
&+ &   \frac{\delta m_{jk}^2}{2 E_\nu} D_s
   \left(1 - \frac{D_d}{D_{ds}} \theta_{\rm{E}}^2 \right) \nonumber \\
&- &  \frac{1}{2 E_\nu} \frac{D_d D_s}{D_{ds}}
   \frac{1}{4}
\Biggl[\delta m_{jk}^2 (\beta^2 + 2 \theta_{\rm{E}}^2)\nonumber \\
& &    \qquad  \qquad \qquad \qquad   + (m_j^2 + m_k^2) \beta \sqrt{\beta^2 
+ 4 \theta_{\rm{E}}^2} \Biggr] . \nonumber \\
\end{eqnarray}
Thus the phase difference is expressed in terms of 
essntially independent astronomical variables: the line-of-sight
distances between observer, deflector and source, the mass of the 
deflector (encoded uniquely in $\theta_{\rm{E}}$ once the distances
have been chosen) and the perpendicular source offset, $\beta$.

Most of the important results obtained towards the end of 
\S\ref{section_nulensing}
can be recast similarly in terms of standard lensing variables, either in
terms of the unobservable image positions or the source position. 
Assuming separability (see \S\ref{section_nulensing}), 
for instance, the 
contribution to the $\Delta \Phi$ due to path difference effects alone
(Eq.~\ref{eqn_phasediff_pq})
can be written as
\bea
\label{eqn_phasediff_pq_dimless}
\Delta \Phi^{pq} &\simeq&
E_\nu \,
\frac{D_d D_s}{D_{ds}} \left[\frac{\Delta \theta^2_{pq}}{2} 
+ \theta_{\rm{E}}^2 \,\ln 
  \left(\left|\frac{\theta_p}{\theta_q}\right|\right) 
\right] \nonumber \\
&-&  \frac{\bar{m}^2}{2 E_\nu} \frac{D_d D_s}{D_{ds}} 
\frac{\Delta \theta^2_{pq}}{2}
\eea
which for $p = +$ and $q = -$ becomes
\begin{eqnarray}
\label{eqn_phasediff_+-_dimless}
\Delta \Phi^{+-} & \simeq &
E_\nu
  \frac{D_d D_s}{D_{ds}} 
  \Biggl[\frac{\beta \sqrt{\beta^2 + 4 \theta_{\rm{E}}^2}}{2} \nonumber \\
&& \qquad \qquad \qquad + \theta_{\rm{E}}^2 \,\ln
  \left(\left|\frac{\beta + \sqrt{\beta^2 + 4 \theta_{\rm{E}}^2}}
  {\beta - \sqrt{\beta^2 + 4 \theta_{\rm{E}}^2}}\right|\right)
\Biggr] \nonumber \\
& & 
- \frac{\bar{m}^2}{2 E_\nu} \frac{D_d D_s}{D_{ds}}
\frac{\beta \sqrt{\beta^2 + 4 \theta_{\rm{E}}^2}}{2}.
\end{eqnarray}
Similarly, the contribution due solely to the different phase velocities
of two mass eigenstates traveling down the same path 
(Eq.~\ref{eqn_phasediff_jk})
can be expressed as
\be
\label{eqn_phasediff_jk_dimless}
\Delta \Phi_{jk} \simeq
  \frac{\delta m_{jk}^2}{2 E_\nu} D_s
    \left[ 1 - \frac{\theta_{\rm{E}}^2}{8} 
      \left( \beta^2 + 4 \theta_{\rm{E}}^2 \right)
    \right].
\ee

\section{The Oscillation `Probability'}
\label{section_osnprob}

With the above results, we can now calculate the analog, in curved space,
 of the flat-space neutrino
oscillation probability:
\be
|\langle \nu_\beta | \nu_\alpha, D_s \rangle|^2 
\propto
\int \mathrm{d}T \vert \langle \nu_\beta \vert \nu_\alpha; A, B) \rangle \vert^2
\ee
so that we can write
\begin{multline}
\label{eqn_oscnprobcurved}
|\langle \nu_\beta | \nu_\alpha, D_s \rangle|^2 \\
= \vert N \vert^2 \sum_{pq} \sqrt{I_pI_q} \sum_{jk} 
U_{\alpha j} U_{\beta j}^* U_{\beta k} U_{\alpha k}^* \\ 
\times \exp\left[-\mathrm{i}\Delta\Phi_{jk}^{pq}(D_s)\right],
\end{multline}
where $I_p$ and $I_q$
account for the fact that different paths may be differentially magnified
by a lens. 
We remind the reader that $|\langle \nu_\beta | \nu_\alpha, D_s \rangle|^2 $
is no longer strictly a probability -- see \S\ref{section_beamsplitter}.

In the case that the `separability' defined by Eq.~(\ref{eqnseparable})
is satisfied, if, for the moment, we are 
interested only in determining the (energy) spacing  of the
interference maxima and minima, we need only consider a 
plane-wave-like calculation (and can
therefore set to one side the coherence length effects and so on 
that emerge from a wavepacket calculation). So, following
considerations similar to those
that lead to Eq.~(\ref{eqn_app_bs''''}) in the appendix we can calculate that
\bmu
\label{gravoscnprobcontd'}
 |\langle \nu_\beta | \nu_\alpha, D_s \rangle|^2  \\
 \shoveleft{
= |N|^2 \left[N_\mathrm{path}I 
+ 2 \sum_{p,q < p} \sqrt{I_p I_q} \cos(\Delta \Phi^{pq})\right]}
\\ \times 
 \biggl[\delta_{\alpha \beta} - 
4Re\biggl\{\sum_{j, k < j}
U_{\alpha j}  U_{\beta j}^* U_{\alpha k}^*  U_{\beta k}
\left[\sin^2\biggl(\frac{\Delta \Phi_{jk}}{2}\right) \\
+ \frac{\mathrm{i}}{2}\sin(\Delta \Phi_{jk}) \biggr] \biggr\}\biggr],
\end{multline}
where $N_\mathrm{path}$ is again the number of classical paths from source to 
detector (two in the case of a Schwarzschild metric), 
$I \equiv \sum_p I_p/N_\mathrm{path}$,  and  
the normalization, $|N|^2$ is again given 
(cf.\ \S\ref{section_beamsplitter}) by requiring that
$\sum_\beta |\langle \nu_\beta | \nu_\alpha, D_s \rangle|^2 \leq 1$, i.e.,
\begin{equation}
max\left\{\langle \nu_\alpha, B | \nu_\alpha, B \rangle\right\} = 1.
\end{equation}
This means that 
\begin{equation}
|N|^2 = \frac{1}{N_\mathrm{path} I + 2\sum_{p,q < p}\sqrt{I_p I_q}}.
\end{equation}
Eq.~(\ref{gravoscnprobcontd'})
establishes 
the contention made above
that 
interference effects that emerge with 
gravitationally-lensed neutrinos 
are a combination of a Young's double slit type interference
\{the $[...\cos(\Delta \Phi^{pq})...]$ envelope term\} and flat space
oscillations [the $(\delta_{\alpha \beta} -...)$ term].
Further, assuming the separability requirement is satisfied, we can see
how $|\langle \nu_\beta | \nu_\alpha, D_s \rangle|^2$ factorises
into an interference pattern and a conditional probability.
This is a repeat of the behavior see in 
\S\ref{section_planewavesplitter}.


One should also note that in the particular case of
the Schwarzschild lens under consideration in the last section, the
two (assuming non-perfect alignment) classical paths from source to detector,  
denoted by the subscripts $+$ and $-$,
experience magnifications given by 
[see Eq.~(2.24) of ref.~\cite{Schneider1992}]
\be
I_\pm 
=\frac{1}{4}\left(\frac{\beta}{\sqrt{\beta^2 +4\theta_\textrm{E}^2}}
+\frac{\sqrt{\beta^2 +4\theta_\textrm{E}^2}}{\beta} \pm 2\right).
\ee
This gives us that
\begin{equation}
|N|^2 = \frac{1}{I_+ + I_- + 2\sqrt{I_+ I_-}} 
= \frac{\beta}{\sqrt{\beta^2 + 4\theta_\textrm{E}^2}}
\end{equation}
Eq.~(\ref{gravoscnprobcontd'}) then becomes:
\bmu
\label{gravoscnprobcontd''}
 |\langle \nu_\beta | \nu_\alpha, D_s \rangle|^2  \\
  \simeq  \frac{1}{\beta^2 + 4\theta_\textrm{E}^2} 
\bigl\{\beta^2 + 2\theta_\textrm{E}^2[1 + 2\cos(\Delta \Phi^{+-})]\bigr\}
\\
\times  \biggl[\delta_{\alpha \beta} - 
4Re\biggl\{\sum_{j, k < j}
U_{\alpha j}  U_{\beta j}^* U_{\alpha k}^*  U_{\beta k}
\left[\sin^2\biggl(\frac{\Delta \Phi_{jk}}{2}\right) \\
+\frac{\mathrm{i}}{2}\sin(\Delta \Phi_{jk}) \biggr] \biggr\}\biggr],
\end{multline}
with $\Delta \Phi^{+-}$ and $\Delta \Phi_{jk}$ given
by Eqs.~(\ref{eqn_phasediff_+-_dimless}) and 
(\ref{eqn_phasediff_jk_dimless}) respectively (where, again, care should 
be taken not to confuse $\beta$ as a label on the neutrino flavor 
with $\beta$ as the source angular position).

Another result of interest is that for the {\it magnification}, 
$\mu_{\nu_\beta}$,
which
is defined to be the ratio of the flux of neutrinos of type
$\beta$ actually received
(from the source at $Y$ and given the lensing mass  is where it is) {\it to}
the flux of
neutrinos of the same type that would be received  with
the lens absent (but with the source in the same position):
\be
\mu_ {\nu_\beta}= \frac{F(D_s, E_\nu) \sum_\alpha 
|\langle \nu_\beta | \nu_\alpha, D_s \rangle|^2_\textrm{lens}
\times P_\alpha}
{F(D_s, E_\nu) \sum_\alpha 
|\langle \nu_\beta | \nu_\alpha, D_s \rangle|^2_\textrm{no lens}
\times P_\alpha},
\ee
where $F(D_s, E_\nu)$ denotes the flux of neutrinos of all types that would be received,
at an energy of $E_\nu$ and factoring in geometrical effects, in the absence of the lens. Also,
$P_\alpha$ denotes the probability that a neutrino
generated by the source under consideration is of type $\alpha$.
Now, for the Schwarzschild lens, assuming mass degeneracy,
the path difference and phase velocity contributions to 
$|\langle \nu_\beta | \nu_\alpha, D_s \rangle|^2$ factorise
into an interference pattern and a conditional probability, as mentioned above (Eq.~(\ref{gravoscnprobcontd'})).
This has the effect  that the magnification
is independent of the neutrino flavor under consideration:
\bea
\label{eqn_magnification}
\mu_{\nu_\beta} &\simeq
\frac{\sum_\alpha |N|^2 \left[I_+ + I_- 
+ 2 \sqrt{I_+ I_-} \cos(\Delta \Phi^{+-})\right]
P(\alpha \to \beta)_\textrm{flat}
\times P_\alpha}
{\sum_{\alpha'} |N|^2  
P(\alpha' \to \beta)_\textrm{flat}
\times P_{\alpha'}}\nn \\
&= \left[I_+ + I_- 
+ 2 \sqrt{I_+ I_-} \cos(\Delta \Phi^{+-})\right] \nn \\
&= \frac{1}{\beta \sqrt{\beta^2 + 4\theta_\textrm{E}^2}}
\bigl\{\beta^2 + 2\theta_\textrm{E}^2[1 + \cos(\Delta \Phi^{+-})]\bigr\},
\eea
where $P(\alpha \to \beta)_\textrm{flat}$, 
the flat space neutrino oscillation probability,
is given by
\bea
P(\alpha \to \beta)_\textrm{flat}
&=&  
\biggl[\delta_{\alpha \beta} \,
- 
4Re\biggl\{\sum_{j, k < j}
U_{\alpha j}  U_{\beta j}^* U_{\alpha k}^*  U_{\beta k} \nn \\
& &\times
\left[\sin^2\biggl(\frac{\Delta \Phi_{jk}}{2}\right) 
+ \frac{\mathrm{i}}{2}\sin(\Delta \Phi_{jk}) \biggr] \biggr\}\biggr]. \nn \\
\eea
The result for the magnification is as expected given what is known about the photon case
[see Eq.~(9) of Ref.~\citep{Stanek1993}].

Finally for this section, we determine, for future reference,
 the fringe {\it visibility}, ${\cal V}(\beta)$:
\bea
\label{eqn_visibility}
{\cal V}(\beta) & 
\equiv & \frac{\mu_{\nu_\beta}^\textrm{max} - \mu_{\nu_\beta}^\textrm{min}}
{\mu_{\nu_\beta}^\textrm{max} + \mu_{\nu_\beta}^\textrm{min}} \nn \\
& \simeq & \frac{2\theta_\textrm{E}^2}{\beta^2 + 2\theta_\textrm{E}^2}. 
\eea



\section{Phenomenology: Heuristic Considerations}
\label{section_phenomenology}

Above we have presented the calculation of the phase and the consequent phase difference,
oscillation probability analog -- $|\langle \nu_\beta | \nu_\alpha, D_s \rangle|^2$ -- 
which determines the form of the oscillation pattern seen at a detector, and
magnification factor, all for the Schwarzschild lens. 
We now turn briefly to the question of the phenomenological consequences of all these 
theoretical developments. We shall deal with the issues presented here at
greater length in another work \citep{Crocker2003}.
There are a number of factors which broadly determine the visibility of GINI effects
\footnote{Note that we will assume separability of the phase difference is satisfied in the
following discussion.}:

\begin{enumerate}
\item \label{point_suitablesources} 
{\bf Suitability of potentially-lensed sources.}
The first consideration must be, what qualifies as a suitable source? We require sources
that produce a neutrino signal that might be both gravitationally
lensed and of sufficient intensity. 
\item \label{point_geometrical_opt} {\bf Geometrical optics limit.}
Because our theoretical evaluation for the neutrino phase difference has been performed within 
the geometrical optics limit (where only the classical paths
from source to detector need be considered in determing the 
form of the interference pattern), we 
require that this limit holds in the experimental situation under consideration.
This translates to the requirement that the de Broglie wavelength of the
neutrino mass eigenstates in not larger than the Schwarzschild radius of the lens (the latter
quantity setting the scale of the path difference: see below)
\citep{Stanek1993,Deguchi1986a,Peterson1991}
\footnote{This is analogous to the 
requirement that, for the `usual' equation (obtained in the geometrical optics limit)
describing the intensity
on a screen in a Young's slit type experiment to be correct,
the wavelength of the interferring radiation must not be larger than the slit separation.
Otherwise, the equation suggests a (non-physical) violation of conservation of total intensity
in the form of
an interference maximum over the whole screen.}.
\item \label{point_detectorEresoln} {\bf Detector energy resolution.} 
Even if there
exists an interference pattern to be mapped out -- and sufficient events to achieve this -- 
a separate question is whether the smearing of this  pattern
caused by the finite energy resolution of any real neutrino detector 
is so large as to completely wash it out. 
\item \label{point_Goldilocks}{\bf Just-so condition for lensing mass.} 
Points \ref{point_geometrical_opt} and \ref{point_detectorEresoln} 
imply a range for 
a `just-right' lensing mass  (given the energy scale of the neutrinos is already set)
-- not too large and not too small --
inside which GINI effects may become evident.
This can be roughly determined by the following
considerations: for a (point mass) lensing system
to produce images of similar brightness (so that
interference effects might be seen),
we require that source be sufficiently well aligned with the line
from the observer through the lens (i.e., $\beta \lesssim \theta_\textrm{E}$).
Granted this, 
 the scale of the path difference is then set by the 
Schwarzschild radius (see, e.g., \cite{Schneider1992}, p. 240), 
\bea
R_S \equiv 2GM \simeq 3 
\times 10^{-12} \left(\frac{M}{10^{-17}\, \msun  } \right) \, \mathrm{cm},\nn \\ 
\eea
of the lens (and -- very broadly -- can be considered as independent from
the distance to the detector), once one has settled on a generic astrophysical source
which emits neutrinos in some characteristic energy range, the lensing mass range is 
determined. This is because
we require
\be
\label{eqn_goldilocks}
E_\nu \, \times \,R_S \simeq 1
\ee 
at an energy either within or not too far below
that detectable by 
the particular
detector technology under consideration 
(see \S\ref{section_energyranges} below). 
We label this constraint on the lensing mass range
the just-so condition.
\item {\bf Wave packet spreading and decoherence}. 
By analogy with the considerations set out in 
\S\ref{section_beamsplitter}, we expect that the full expression
for the oscillation probability analog include
exponential decay factors that account for coherence loss effects.
These essentially factor in
the interference attenuation which occurs when
the different neutrino wavepackets, traveling
with different group velocities 
and/or along paths of different affine length,
overlap significantly less than completely at the detector.
See \S\ref{section_curved_space_decoherence} for more detail on this
issue.
\item {\bf Finite source  size effects.} Our derivation of the phase difference
has assumed a stationary point source (and detector). 
Of course, this is at variance with Heisenberg uncertainty requirements. But
more significantly, any real, {\it macroscopic} source (the region giving birth to 
all the neutrinos that
are identified as having come from a particular astrophysical object) will be of
finite -- indeed macroscopic -- size. This can, like detector
energy resolution issues, tend to 
wash away the interference pattern because the path difference is now different for the various
neutrinos that come from different parts of the `same' object. More concisely, an effective
source angular extent 
of the order of -- or larger than -- the angular extent of the Einstein radius
means that the visibility of the interference fringes is reduced \citep{Gould1992}. 
If the source size
 is denoted by $r_\textrm{source}$, 
then
this translates to the requirement that
\be
\label{eqn_pointsourcecondn}
r_\textrm{source} < \sqrt{4 GM \frac{D_s D_{ds}}{D_d}}.
\ee
Eq.~(\ref{eqn_pointsourcecondn}) is not a sufficient
condition to guarantee a point-like source, however;  
as  energy -- and, therefore, phase
along any particular path -- increases, there will come a point where 
(while the lens-induced path difference might still generate the greatest component of the
phase difference for neutrinos from all parts of the lens)
the phase difference for neutrinos emerging from one part of the source
will be noticably different to that for neutrinos generated
from a different part of the source.
At this point the interference pattern will, again, become
smeared out.
That this {\it not} occur bounds the energy:
\be
\label{eqn_strictpointsourcecondn'}
E_\nu \lesssim \frac{2 D_s D_{ds}}{D_d \,\, r_\mathrm{ source }^2 }\,.
\ee
\item {\bf Finite detector size effects.} Much of the 
discussion immediately above carries through,
{\it mutatis mutandis}, to considerations stemming from finite detector size.
Explicitly, finite detector size effects can tend to wash away
the interference pattern because the path difference (at any particular energy)
will be non-constant across the volume
of the detector. One must determine whether this is a significant effect.
\item {\bf Finite lens size effects.}
We have calculated the neutrino phase difference in a Schwarzschild metric, i.e., assuming
the lens to be effectively pointlike.
This assumption will hold, at least roughly, 
if the Einstein radius
of the lensing system 
[$\theta_E$, as defined in Eq.~(\ref{equation:theta_e})]
is larger than the scale of the physical dimension of the lens. 
\item {\bf Source-lens-alignment probability.} In order 
to see interference fringes we require that the
{\it visibility} [defined in Eq.~(\ref{eqn_visibility})] be sufficiently good. 
This requires
a sufficient degree of alignment between source, lens, and detector 
(i.e., a small $Y$ or $\eta$).
One can then ask, 
given the lensing mass scale, as determined by point 
\ref{point_Goldilocks} above, and the
expected distance to a source (of the chosen, generic type),
how likely is it that there is a lens within a certain 
distance of the line from the source to the detector?
\item {\bf Time scale of lens crossing.} 
Further to the point immediately above, 
one must consider 
over what time scale the lens will cross the `beam' from 
source to detector and, therefore, how temporally-stable 
-- and, indeed, how long-lasting -- 
any interference pattern will be.
\item {\bf Intrinsic Source Spectrum.} In order to confidently identify
interference effects one must be able to rule out the possibility of
the intrinsic spectrum of the source mimicking these effects. 
Moreover, even given
a well-understood source spectrum, a separate question is whether there 
is a measurable neutrino
flux over a sufficient energy extent that a number of interference fringes 
might be seen at a detector. 
\end{enumerate}

\subsection{More Detail on Decoherence Effects in Curved Spacetime}
\label{section_curved_space_decoherence}

We only attempt an heuristic treatment 
here\footnote{see the Appendix of Ref.~\cite{Cardall1997}
for a treatment of the direct analog of the coherence length 
of neutrino oscillations in flat space for 
curved spacetime,
though note that the treatment presented here does not apply for multiple,
macroscopic paths.}.  
Ignoring detector effects (see below), coherence requires that  there is
significant overlap between \MES s at a detector. 
As explained in the appendix, the various \MES s,
 may have traveled with both different
group velocities and along different paths.
Let us take a source located on a source plane at $D_s$ and
neutrino \MES s with an effective width  of $\sigma_x$.  
Then, by analogy with
the second exponential damping term in Eq.~(\ref{eqn_app_bs''''}) of the
appendix and 
given the scale of the path difference is given by $R_S$, 
interference between \MES \ $j$ traveling
down one macroscopic path through a Schwarzschild spacetime and 
$k$ the other, roughly requires:
\be
\label{eqn_curvedcoherence}
\left(R_S \mp \frac{\vert \delta m^2_{jk}\vert}{2 E_\nu^2} 
D_s \right)^2 \lesssim 8 \sigma^2_x.
\ee
Here the upper sign refers to the case when the lighter 
\MES \ traverses the longer path,
the heavier along the shorter path,
and the lower sign refers to the opposite case (there are now four broad 
cases depending on this sign
and the relative sizes of $R_S$ and $2 \sqrt{2} \sigma_x$).

Note that if we wish to consider interference between different \MES s 
traveling down the {\it same}
path -- i.e., the direct analog of flat space neutrino oscillations -- we take $R_S \to 0$ in 
in Eq.~(\ref{eqn_curvedcoherence})
 so that we require
\be
\label{eqn_nuoscncoh}
D_s \lesssim \frac{2 E_\nu^2}{\vert\delta m^2_{jk}\vert}2 \sqrt{2} \sigma_x,
\ee
(then the equality in the above is satisfied for 
$D_s \simeq L_\textrm{coh}$, where $L_\textrm{coh}$ is the coherence length),
whereas if we wish to consider interference between the same \MES \ traveling
down different paths, then from Eq.~(\ref{eqn_curvedcoherence}) we require
\be
R_S \lesssim 2 \sqrt{2} \sigma_x.
\ee
Below we shall determine some plausible numbers to put in these relations 
(for a number of different
neutrino sources), 
but first we recall some considerations behind the determination of $\sigma_x$.



\subsection{Determining the Size of the Wavepacket}
\label{section_wvepktsize}

At an heuristic level 
-- adequate to the order of magnitude calculations we will make -- 
the neutrino wavepacket size (in position space) is given by the size, 
$d$, of the
region to which the neutrino parent particle is localized 
\citep{Kayser1981,Kim1993}\footnote{Note that we ignore here 
the contribution of the {\it detection} 
process to the effective wavepacket size that can, 
in principle, act to restore coherence 
via broadening the effective wavepacket width through accurate energy/momentum
measurement:
see Ref.~\citep{Giunti1998a} for more detail here and 
also  Ref.~\citep{Beuthe2002} 
and for a rigorous, quantum-field-theoretic treatment of neutrino coherence 
length.}:
\be
\sigma_x \qquad \simeq \qquad d.
\ee
In turn, $d$ is related to $t_\mathrm{eff}$, the effective time available for
the coherent emission, by the parent, 
of a neutrino wave train:
\be
d \simeq t_\mathrm{eff} \,.
\ee
In free space the coherent emission time corresponds to the decay 
time of the parent particle, $\tau$, but if the parent particle is in 
a dense and hot medium and undergoing 
collisions with its neighbors on a timescale, $t_\mathrm{collision}$, 
smaller than $\tau$, 
then $t_\mathrm{eff} \simeq t_\mathrm{collision}$ 
\citep{Nussinov1976,Anada1988}. 
This effect corresponds to the collision or pressure broadening of
atomic spectral lines. 
In summary, we shall take
\begin{equation}\label{sigxfromtau}
\sigma_x  \simeq  \gamma \, t_\mathrm{eff} 
\qquad \mathrm{with} \qquad t_\mathrm{eff} = min\{t_\mathrm{collision}, \tau \},
\end{equation}
where we have explicitly introduced a Lorentz boost, $\gamma$,
to allow for any bulk motion of the source with respect to the detector frame.
This factor can, of course, be large for astrophysical sources.

\subsection{Determing Energy Ranges for GINI Phenomenology}
\label{section_energyranges}

There are two energy ranges that must be considered in our analysis, viz
\begin{enumerate}
\item {\bf Extrinsic energy range.}
Forgetting GINI effects for the moment, one energy range -- which we label extrinsic -- 
is delimited
by the minimum and maximum energies, $E_\textrm{min}$ and  $E_\textrm{max}$,
at which the generic source under consideration
can be seen in neutrinos by a particular detector
technology.
These limiting energies are determined by either detector or intrinsic source spectrum
considerations (whichever is the more severe). 
The extrinsic energy range is defined by
\be
E_\textrm{min} \lesssim E_\nu \lesssim E_\textrm{max}.
\ee

\item {\bf Intrinsic energy range.}
We also identify an intrinsic energy range that is given by the following considerations: the
lower bound on this range is given by the critical energy, $E_\textrm{crit}$,
 at which the pertinent phase 
difference 
is equal to one (below this value our treatment of the phase breaks down). The scale of
this energy is given by requiring
\be
E_\textrm{crit} \simeq \frac{\hbar c}{R_S}.
\ee
Note that the relation is not exact because the RHS does not account for
the effect of the source alignment parameter, $Y$, on the phase difference.
The upper bound on this range, $E_\textrm{washout}$,
is determined by the energy at which detector energy resolution issues mean that
one interference fringe can no longer be resolved from another. 
Washout occurs generically because, although interference fringes are distributed
at equal energy intervals, the absolute uncertainty in neutrino energy determined by a
detector can be expected to be an increasing function of energy. 
\end{enumerate}

\section{Suitable Source - Lens - Detector Configurations for GINI}

We can think of four scenarios for source - lens - detector configuration that {\it might} 
exhibit
GINI effects (there may well be more), viz:
\begin{enumerate}
\item Sun -- Moon -- solar neutrino detector (i.e., in an Solar eclipse)
\item cosmological neutrino source -- intervening lensing object 
-- large scale Water/Ice {\v C}erenkov neutrino
detector or airshower array
\item artificial neutrino beam on one side of earth aimed through center of earth to 
detector on opposite side of the planet
\item Galactic (i.e., Milky Way) Core Collapse supernova (Types II, Ib and Ic) -- 
intervening lensing 
object -- solar neutrino detector
\end{enumerate}
Unfortunately, scenarios 1. to 3. fail one or more of the 
heuristic tests we have set out above and we must, reluctantly, dismiss them. 
Scenario 4, however, holds out some
promise and it is to this that we now briefly turn 
(see \citep{Crocker2003} for more detail on all the scenarios mentioned), though
we alert the reader from the beginning that scenario is unlikely to be realised 
{\it at present} because of the low probability of supernovae at (neutrino-)detectable
distances being lensed by objects in suitable mass range.

\subsection{Core Collapse Supernovae as Sources for GINI Observation}
\label{section_SN}

\subsubsection{Core Collapse Supernovae as Neutrino Sources: General Considerations}

Let us take the characteristic scale of the distance to a Galactic core collapse supernova to be
10 kpc $\simeq 3 \times 10^{22}$ cm, 
the approximate distance to the Galactic Center.
A core collapse SN observed today at the fiducial 10 kpc 
 would produce 
around $10^4$ 
and $10^3$ 
events in SuperKamiokande 
and the Sudbury Neutrino Observatory respectively 
\citep{Beacom1999}. 
Over the medium term, prospects for SN neutrino detection may 
become even better than at present
with the construction of the next generation of 1 Mt 
underground, water {\v C}erenkov detectors
\citep{Nakaya2002a,Antonioli1999,Jung2000}.
For a 
supernova at  10 kpc, a 1 Mt device
should detect ${\cal O}(10^5)$ events \citep{Jung2000}.

\subsubsection{Natural Scale for Lensing Mass Required for GINI Effects with SN Neutrinos}

 Writing
\bea
\Delta \Phi^{+-} 
&\sim& E_\nu R_S \nn \\
&\simeq& 1.5 \times 10^{17} \left(\frac{E_\nu}{10 \, \textrm{MeV}}\right) 
\left(\frac{M}{\msun}\right) ,
\eea
we can determine that the {\it smallest} lensing mass
that might produce a phenomenological effect (that we can treat using our formalism) is, 
very roughly, $10^{-17} \, \msun \, \simeq \, 3 \times 10^{16}$ g.
This is in the cometary mass range. 
A more detailed calculation \citep{Crocker2003}
demonstrates -- for the specific case of SuperKamiokande -- a
sensitivity to the GINI effect with lensing masses in the range
\bea
\label{eqn_massrange}
10^{-18} \, \msun \lesssim M_\textrm{lens} \lesssim 10^{-16} \, \msun.
\eea
This range is is both conservative and fairly
sensitive to the SuperK energy thresholds and energy resolution. 

\subsubsection{Coherence of Supernova Neutrinos}
A
 neutrino wavepacket
leaving the neutrinosphere of a nascent neutron star will have a size \citep{Anada1990,Kim1993}
\be
\sigma_x^{SN} \simeq 10^{-9} \, \textrm{cm}.
\ee
This is to be contrasted with the scale of the affine path difference for the 
lensing mass range under
consideration (Eq.~(\ref{eqn_massrange})) which is supplied by the range of 
the Schwarzschild radius,
viz:
\bea
3 \times 10^{-13} \cm \gtrsim &R_S& \gtrsim 3 \times 10^{-11} \cm.
\eea
We do not, therefore, expect any significant damping of the interference amplitude by
decoherence due to path difference effects.

There is, however, 
also decoherence due to group velocity difference to be considered, i.e.,
the direct analog of decoherence effects for conventional neutrino oscillations. The
inequality to be satisfied is given by Eq.~(\ref{eqn_nuoscncoh}), 
the RHS of which translates to $\sim 3 \times 10^{13}$ cm for 10 MeV neutrinos
\footnote{We take the {\it largest} possible value for this quantity by assuming 
the scale of the smallest 
experimentally-determined $\delta m^2$, i.e., the solar mass splitting at 
${\cal O} (10^{-5})$ eV$^2$.}
much smaller than the fiducial scale of $D_s$,  $\sim 3 \times 10^{22}$ cm. 
We can expect, therefore, to be
beyond the flat space coherence length. This means that
the neutrino signal will be characterised by flavor ratios that are
constant across (measurable) energy. 
For supernova
neutrinos, then, 
if a suitable lens were present, GINI would cause
patterns of maxima and minima across energy 
in the detected neutrino spectra. 
Furthermore,
the positions, in energy, 
of these maxima and minima would be essentially the same
for all
neutrino flavors 
Interference effects
would be, in principle, directly evident even in neutral current interaction data.
On the other hand, we would not expect a noticable change in the ratios between
different neutrino species across energy. In other words, for supernova neutrinos,
given a suitable lens, it is possible to see interference effects due to path
difference effects but not due to phase velocity difference (i.e., flat space
oscillation) effects.

\subsubsection{Finite Source Effects with Supernova Neutrinos}

Given a scale for the neutrinosphere, $r_{SN}$, of $\sim 10$ km $= 10^6$ cm,
a calculation shows that the point source condition, 
Eq.~(\ref{eqn_pointsourcecondn}), 
fails at the lower end of the 
of phenomenologically-interesting lensing mass range {\it assuming} $D_{ds} \simeq D_d$. 
Furthermore, from Eq.~(\ref{eqn_strictpointsourcecondn'})
we find that in order that the phase uncertainty introduced by the finite size of
the  supernova neutrino source not be too large, we require that the neutrino energy 
be less than $\sim 1$ MeV,
a condition that, with 10 MeV neutrinos, we fail to meet by an order of
magnitude, {\it again assuming} $D_{ds} \simeq D_d$.
We hasten to add, however, that 
we do not believe that either of these two is necessarily fatal: a numerical
study is needed here and this may well establish that
GINI effects are visible even when the crude, heuristic inequalities above are violated.
\footnote{Certainly, in their numerical
study of femtolensing  with a disk source, Peterson and Falk \citep{Peterson1991},
found that, allowing for a realistic deviation from smoothness in
the source function -- which describes the intensity across the disk of the
source -- 
  interference effects were visible with a source size  significantly larger than the
Einstein ring. We have not allowed for this (potential) effect for
a supernova neutrino source.}
Moreover, that $D_{ds} \simeq D_d$ need not 
hold (over the Galactic scales we are considering) and, further, we might have
$D_s >$ 10 kpc (at the cost of a reduced event rate). In either case
point source conditions could  easily be satisfied.

\subsubsection{Finite Lens Size Effects}

For a lens in the mass range determined above, and both source and lens
at Galactic length scales, the Einstein length scale is
\be
D_d \theta_E
\simeq 3 \, \sqrt{\left(\frac{M}{10^{-17} \msun}\right)
\left(\frac{D_d}{D_s}\right)
\left(\frac{D_{ds}}{5 \kpc}\right)} \km.
\ee
On the other hand, for a lens with the density of $\sim 1$ gm cm$^{-1}$, 
the scale of the
dimensions of the
lensing object, $l$, is given by 
\be
l \simeq 0.6 \left(\frac{M}{10^{-17} \msun}\right)^{\frac{1}{3}} \km,
\ee
meaning that the classical paths pass very close to the object, and, 
in some cases, one path might actually pass inside the object. 
Given the order of magnitude nature of the calculations we have 
performed here, however, this fact will not significantly impinge 
on the observability of the effect we predict. 
Certainly the neutrinos will not interact significantly
with the material of the lens. 
Of course, if the lensing object is a black hole, taking 
the lens to be a point source is unproblematic.

\subsubsection{Finite Detector Size Effects}

A quick calculation shows \citep{Crocker2003} that
finite detector size/position resolution effects never become insurmountable
over the whole range of possible lens positions.

\subsubsection{Crossing Time Scales for Supernova Neutrinos}

If the lens has a transverse velocity $v \simeq 30 \ \textrm{km \ s}^{-1}$, 
it will move across the Einstein 
ring in a time scale of $\sim 1$ s
\citep{Gould1992}.
Given, then, that we expect a detectable neutrino signal will be received
from a SN over a period of around 10 seconds, 
interference fringes will shift over the time of observation, but not
so quickly that they cannot be observed.

\subsubsection{Lensing Probability}

The above paragraphs detail the conditions that a source-deflector-observer
alignment must satisfy in order for GINI to be measured, but implicit at
all stages is that such an alignment has occured. Unfortunately the 
chance of a suitable deflector lying sufficiently close to the line-of-sight
to a source in the Local Group is not high. 

The Galaxy's rotation curve places a strict upper bound on the total
mass in its halo (e.g., \cite{Sakamoto2003}), which then 
implies a maximum possible
alignment probability to, say, the Magellanic Clouds. Even if the halo
consisted only of point-masses of suitable size, the simple calculation
made by Paczynski \cite{Paczynski1986} 
implies that the lensing optical depth, $\tau$ -- essentially equal
to the probability that any single source is lensed at a given time --
of $\sim 10^{-6}$. This result has been corroborated experimentally by
monitoring stars in both the Galactic center and the Large Magellanic
Cloud for period variations: 
both the MACHO \cite{Alcock2000} and OGLE \cite{Wozniak2001}
groups 
have found $\tau \simeq 3 \times 10^{-6}$. 
It is important, then, to note that even if the halo is dominated by 
point-masses of $M \simeq 10^{-17} \msun$ suitable for GINI, 
the alignment probability 
to any neutrino sources sufficiently close to be detected at all is only
$\sim 10^{-6}$.

In the future, however, as detector technology improves, it may be possible
to observe neutrinos from cosmologically distant sources at effective
distances of up to Gpc. The lensing optical depth is thus increased, 
both because a given mass can act as a more efficient lens and because
the chance of alignment increases proportionally with source distance. 
A simple calculation of these effects (e.g., \cite{Schneider1992}) implies that optical
depths of close to unity are plausible; thus when neutrinos are detected
from cosmologically distant sources GINI effects will {\it have} to be taken
into account in the interpretation of any such data obtained.


\section{Extention of Theoretical Results} 
\label{section_furtherwk}

Besides treating the potential phenomenological effects of GINI at 
greater length in another work \citep{Crocker2003},
there are, of course, a number of directions in which our theoretical
treatment will be extended.
Some issues we intend dealing with further in another publication 
\citep{Crocker2003}
include:
\begin{enumerate}
\item \label{point1}

From consideration of interference of lensed {\it photons}
in a Schwarzschild metric
[see Eq.~(7.8) of \cite{Schneider1992} and also
see Eq.~(9) of \cite{Stanek1993}]
Eq.~(\ref{gravoscnprobcontd''}), 
is actually subtly in error: there is an extra  $-\pi/2$ phase shift
missing from the argument of $\cos$ term (in other words, the 
interference envelope should actually go as 
$\beta^2 + 2\theta_\textrm{E}^2[1 + \sin(\Delta \Phi^{+-})]$, generating a 
central minimum for $Y=0$). This is present in the
case of light -- and will also be present in the case of neutrinos -- 
because of the opposite parities of the two images produced
by a Schwarzschild lens 
(i.e., the images -- {\it were} they able to be distinguished -- 
would be flipped with respect to each
other: \footnote{An experimental analog of interference in such a situation
would be a Young's double 
slit apparatus where one beam of light is reflected in a
mirror. This produces, of course, a central minimum.}). 
The reason why our treatment has failed to pick this extra phase shift up
is that we have artificially restricted the paths under consideration
to only the classical ones. 
In other words, 
we have assumed the geometrical optics limit which is strictly only
valid for
phase differences 
of order unity
and larger.
A more complete treatment using the techniques of physical optics -- involving
integration over {\it all} paths through the lens plane
(each such path being uniquely specified by its impact parameter) -- would
recover this phase [and, in fact, demonstrate that the full expressions
for the oscillation probability and magnification
involve confluent Hypergeometric functions
that only reduce to trigonometric functions in the large phase limit: cf.\
Eq.~(7.11) of Ref.~\citep{Schneider1992}
or Eq.~(6) of \citep{Stanek1993}]. 
Moreover, a more complete treatment would also demonstrate that the  singularity
at $\beta = 0$ for Eq.~(\ref{eqn_magnification}) is not a real effect.
\item \label{point2}
So far we have assumed a static metric. But the GINI effect, as noted,
requires
 neutrinos from astrophysical sources 
that would probably need to be located at extra-galactic or even
cosmological distances 
for a decent chance that lensing occur
(though it should be stressed that
current detector technologies would not allow detection of neutrinos from supernovae
beyond out Galaxy and its satellites: \citep{Beacom1999}) and
the introduction of cosmology into the formalism developed here
would
require the treatment of a non-static metric \citep{Wagner1998,Mbonye2001}.
We note in passing that were GINI effects ever 
seen in neutrinos from cosmological sources, these effects would provide for a  test 
of quantum mechanics over the very longest scales. We speculate, then,
that GINI could be sensitive to the effects of spacetime foam (cf., say,
\citep{Ragazzoni2003}). In principle, we also expect that a GINI
pattern in cosmologically-sourced neutrino could be interrogated to determine the value of
the
Hubble constant H$_0$
\citep{Bolton2003,Refsdal1964}. Observation (or non-observation) of GINI
effects would also contitute a {\it de facto} probe of the distribution
of dark matter
objects within a certain well-defined (and interesting) mass range \citep{Gould1992}.
%
\item \label{point3}
The Schwarzschild lens is an ideal case never precisely encountered in nature. 
For the realistic situation we may need to  account for
shear \citep{Ulmer1995}, multiple lensing masses, etc.
Treating these effects may very well demonstrate
 that 
the just-so mass range is considerably larger 
than the estimate given by Eq.~(\ref{eqn_massrange}) 
\citep{Ulmer1995}. 
In any case, it is fairly easy to understand, at least
at the heuristic level, how more general lenses might be treated: 
considering, say, Eq.~(\ref{nonradschwarzphasediffdimless2})
it can be seen that almost all the contributions to the phase
difference are essentially geometrical. The only expression which
contains information on the mass distribution of the deflector
is the logarithmic term. This suggests that it may be possible to modify our results
to arbitrary weak deflectors simply by inserting the appropriate
lens potential (and replacing a Schwarzschild-specific result
for $\Delta \theta_{pq}^2$).
\item \label{point4} In \S\ref{section_phenomenology} we added into the mix
coherence loss considerations. Formally, these can only arise
in a full wavepacket treatment which we have not attempted for the curved 
spacetime case. 
We remind the reader, however, that 
our results for phase and phase-difference will continue to hold in any more detailed
calculation because these are independent of wavepacket considerations.
\item \label{point5}
Our treatment of effects due to the 
finite nature 
of
any real source is very much at an heuristic level.
 Furthermore, finite detector effects can also be important, as has been remarked.
In this regard,
note, in passing, that logically connected to this concern is the consideration
that the observation of GINI effects with two -- or, preferably, more --
widely-separated detectors holds out some interesting possibilities
\citep{Gould1992}. 
One would expect here that the interference patterns seen by different
detectors be, in general, displaced in energy with respect to each other. The degree of
this displacement will be related to the lensing mass and the geometry.
So the displacement could probably be used
to better
constrain relevant parameters than observation with a single detector. Moreover, observation
of fringes with more than one neutrino detector would certainly lend credence to the idea
that these have their origins in GINI.

\end{enumerate}

\section{Conclusion}
\label{section_conclusion}

In this paper we have  explicitly
calculated the  phase  for a neutrino \MES \ propagating through
curved spacetime, in particular, a Schwarzschild metric.
With this expression in hand, we have shown how a novel
interference effect -- gravitationally-induced neutrino interference (`GINI') -- 
may show up for
gravitationally-lensed, astrophysical neutrinos.
These interference effects lead to a neutrino transition
phenomenology qualitatively different from flat space neutrino
oscillations.
We have shown, further, that
a result extant in the literature \citep{Fornengo1997}
for the phase difference with gravitational lensing must be in error.
We have also derived the form of this phase difference when it is given in 
terms of conventional lensing parameters. We have derived the analog of the
neutrino oscillation probability in flat space for the Schwarzschild metric. 
This quantity controls the phenomenology
at a detector, in particular, the pattern of maxima and minima (across energy)
for neutrino wavepackets which have propagated from source to detector along multiple paths.
We have adduced heuristic arguments that establish that this interference pattern
could be seen in the neutrino signal from a supernova, {\it provided} a suitably-lensed
supernova event occurs.
Current -- and probably even next-generation -- neutrino detector technologies
would seem to mean, however, that the probability of such lensing occurring for
a neutrino-detectable supernova is small. Still, for astrophysical neutrinos
originating at cosmological distances the lensing probability approaches 1 and some
day the technology to detect large numbers of these from single sources may become
available.
We have mapped out a program for further research in this field.

In summary, the
material presented in this paper serves as
a proof-of-principle that the GINI effect is both real -- in a theoretical sense -- and,
what is more, could lead, one day, 
to interesting phenomenological consequences for supernova neutrinos.

\section{Acknowledgements}

R.M.C. would like to sincerely thank Paul Alsing, Jesse Carlsson, Tim Garoni, 
Matt Garbutt, Sasha Ignatiev, Bruce McKellar, Andrew Melatos,
Keith Nugent, Alicia Oshlack, David Paganin, Andrew Peele, 
 Ray Protheroe, Georg Raffelt, Rob Scholten, Cath Trott,
Rachel Webster, and Stuart Wyithe for enlightening 
discussions. 
He also thanks Nicole Bell and John Beacom 
for a useful correspondence.
Finally, he particularly thanks 
German K{\" a}lberman, Ray Volkas,  and  Randall Wayth 
who all devoted considerable time to setting this authour straight
on a number of subtle issues.

\vspace{0.3cm}
\noindent
D.J.M. is supported by PPARC.

\vspace{0.3cm}
\noindent
The authors thank an anonymous referee for making
comments that led to improvements to this work.

\begin{center}
{\it This paper is dedicated to the memory of Professor Geoffrey Opat.}
\end{center}


\begin{thebibliography}{70}
\expandafter\ifx\csname natexlab\endcsname\relax\def\natexlab#1{#1}\fi
\expandafter\ifx\csname bibnamefont\endcsname\relax
  \def\bibnamefont#1{#1}\fi
\expandafter\ifx\csname bibfnamefont\endcsname\relax
  \def\bibfnamefont#1{#1}\fi
\expandafter\ifx\csname citenamefont\endcsname\relax
  \def\citenamefont#1{#1}\fi
\expandafter\ifx\csname url\endcsname\relax
  \def\url#1{\texttt{#1}}\fi
\expandafter\ifx\csname urlprefix\endcsname\relax\def\urlprefix{URL }\fi
\providecommand{\bibinfo}[2]{#2}
\providecommand{\eprint}[2][]{\url{#2}}

\bibitem[{\citenamefont{{Overhauser} and {Colella}}(1974)}]{Overhauser1974}
\bibinfo{author}{\bibfnamefont{A.~W.} \bibnamefont{{Overhauser}}}
  \bibnamefont{and}
  \bibinfo{author}{\bibfnamefont{R.}~\bibnamefont{{Colella}}},
  \bibinfo{journal}{Physical Review Letters} \textbf{\bibinfo{volume}{33}},
  \bibinfo{pages}{1237} (\bibinfo{year}{1974}).

\bibitem[{\citenamefont{{Colella} et~al.}(1975)\citenamefont{{Colella},
  {Overhauser}, and {Werner}}}]{Colella1975}
\bibinfo{author}{\bibfnamefont{R.}~\bibnamefont{{Colella}}},
  \bibinfo{author}{\bibfnamefont{A.~W.} \bibnamefont{{Overhauser}}},
  \bibnamefont{and} \bibinfo{author}{\bibfnamefont{S.~A.}
  \bibnamefont{{Werner}}}, \bibinfo{journal}{Physical Review Letters}
  \textbf{\bibinfo{volume}{34}}, \bibinfo{pages}{1472} (\bibinfo{year}{1975}).

\bibitem[{\citenamefont{{Greenberger} and
  {Overhauser}}(1979)}]{Greenberger1979}
\bibinfo{author}{\bibfnamefont{D.~M.} \bibnamefont{{Greenberger}}}
  \bibnamefont{and} \bibinfo{author}{\bibfnamefont{A.~W.}
  \bibnamefont{{Overhauser}}}, \bibinfo{journal}{Reviews of Modern Physics}
  \textbf{\bibinfo{volume}{51}}, \bibinfo{pages}{43} (\bibinfo{year}{1979}).

\bibitem[{\citenamefont{{Stanek} et~al.}(1993)\citenamefont{{Stanek},
  {Paczynski}, and {Goodman}}}]{Stanek1993}
\bibinfo{author}{\bibfnamefont{K.~Z.} \bibnamefont{{Stanek}}},
  \bibinfo{author}{\bibfnamefont{B.}~\bibnamefont{{Paczynski}}},
  \bibnamefont{and}
  \bibinfo{author}{\bibfnamefont{J.}~\bibnamefont{{Goodman}}},
  \bibinfo{journal}{ApJL} \textbf{\bibinfo{volume}{413}}, \bibinfo{pages}{L7}
  (\bibinfo{year}{1993}).

\bibitem[{\citenamefont{{Schneider} and {Schmid-Burgk}}(1985)}]{Schneider1985}
\bibinfo{author}{\bibfnamefont{P.}~\bibnamefont{{Schneider}}} \bibnamefont{and}
  \bibinfo{author}{\bibfnamefont{J.}~\bibnamefont{{Schmid-Burgk}}},
  \bibinfo{journal}{A\&A} \textbf{\bibinfo{volume}{148}}, \bibinfo{pages}{369}
  (\bibinfo{year}{1985}).

\bibitem[{\citenamefont{{Peterson} and {Falk}}(1991)}]{Peterson1991}
\bibinfo{author}{\bibfnamefont{J.~B.} \bibnamefont{{Peterson}}}
  \bibnamefont{and} \bibinfo{author}{\bibfnamefont{T.}~\bibnamefont{{Falk}}},
  \bibinfo{journal}{ApJL} \textbf{\bibinfo{volume}{374}}, \bibinfo{pages}{L5}
  (\bibinfo{year}{1991}).

\bibitem[{\citenamefont{{Gould}}(1992)}]{Gould1992}
\bibinfo{author}{\bibfnamefont{A.}~\bibnamefont{{Gould}}},
  \bibinfo{journal}{ApJL} \textbf{\bibinfo{volume}{386}}, \bibinfo{pages}{L5}
  (\bibinfo{year}{1992}).

\bibitem[{\citenamefont{{Ulmer} and {Goodman}}(1995)}]{Ulmer1995}
\bibinfo{author}{\bibfnamefont{A.}~\bibnamefont{{Ulmer}}} \bibnamefont{and}
  \bibinfo{author}{\bibfnamefont{J.}~\bibnamefont{{Goodman}}},
  \bibinfo{journal}{ApJ} \textbf{\bibinfo{volume}{442}}, \bibinfo{pages}{67}
  (\bibinfo{year}{1995}).

\bibitem[{\citenamefont{{Mandzhos}}(1981)}]{Mandzhos1981}
\bibinfo{author}{\bibfnamefont{A.~V.} \bibnamefont{{Mandzhos}}},
  \bibinfo{journal}{Soviet Astronomy Letters} \textbf{\bibinfo{volume}{7}},
  \bibinfo{pages}{213} (\bibinfo{year}{1981}).

\bibitem[{\citenamefont{{Ohanian}}(1983)}]{Ohanian1983}
\bibinfo{author}{\bibfnamefont{H.~C.} \bibnamefont{{Ohanian}}},
  \bibinfo{journal}{ApJ} \textbf{\bibinfo{volume}{271}}, \bibinfo{pages}{551}
  (\bibinfo{year}{1983}).

\bibitem[{\citenamefont{{Deguchi} and
  {Watson}}(1986{\natexlab{a}})}]{Deguchi1986a}
\bibinfo{author}{\bibfnamefont{S.}~\bibnamefont{{Deguchi}}} \bibnamefont{and}
  \bibinfo{author}{\bibfnamefont{W.~D.} \bibnamefont{{Watson}}},
  \bibinfo{journal}{ApJ} \textbf{\bibinfo{volume}{307}}, \bibinfo{pages}{30}
  (\bibinfo{year}{1986}{\natexlab{a}}).

\bibitem[{\citenamefont{{Deguchi} and
  {Watson}}(1986{\natexlab{b}})}]{Deguchi1986b}
\bibinfo{author}{\bibfnamefont{S.}~\bibnamefont{{Deguchi}}} \bibnamefont{and}
  \bibinfo{author}{\bibfnamefont{W.~D.} \bibnamefont{{Watson}}},
  \bibinfo{journal}{\prd} \textbf{\bibinfo{volume}{34}}, \bibinfo{pages}{1708}
  (\bibinfo{year}{1986}{\natexlab{b}}).

\bibitem[{\citenamefont{{Hirata} et~al.}(1987)\citenamefont{{Hirata}, {Kajita},
  {Koshiba}, {Nakahata}, and {Oyama}}}]{Hirata1987}
\bibinfo{author}{\bibfnamefont{K.}~\bibnamefont{{Hirata}}},
  \bibinfo{author}{\bibfnamefont{T.}~\bibnamefont{{Kajita}}},
  \bibinfo{author}{\bibfnamefont{M.}~\bibnamefont{{Koshiba}}},
  \bibinfo{author}{\bibfnamefont{M.}~\bibnamefont{{Nakahata}}},
  \bibnamefont{and} \bibinfo{author}{\bibfnamefont{Y.}~\bibnamefont{{Oyama}}},
  \bibinfo{journal}{Physical Review Letters} \textbf{\bibinfo{volume}{58}},
  \bibinfo{pages}{1490} (\bibinfo{year}{1987}).

\bibitem[{\citenamefont{{Bionta} et~al.}(1987)\citenamefont{{Bionta},
  {Blewitt}, {Bratton}, {Caspere}, and {Ciocio}}}]{Bionta1987}
\bibinfo{author}{\bibfnamefont{R.~M.} \bibnamefont{{Bionta}}},
  \bibinfo{author}{\bibfnamefont{G.}~\bibnamefont{{Blewitt}}},
  \bibinfo{author}{\bibfnamefont{C.~B.} \bibnamefont{{Bratton}}},
  \bibinfo{author}{\bibfnamefont{D.}~\bibnamefont{{Caspere}}},
  \bibnamefont{and} \bibinfo{author}{\bibfnamefont{A.}~\bibnamefont{{Ciocio}}},
  \bibinfo{journal}{Physical Review Letters} \textbf{\bibinfo{volume}{58}},
  \bibinfo{pages}{1494} (\bibinfo{year}{1987}).

\bibitem[{\citenamefont{{Bahcall} and {Glashow}}(1987)}]{Bahcall1987}
\bibinfo{author}{\bibfnamefont{J.~N.} \bibnamefont{{Bahcall}}}
  \bibnamefont{and} \bibinfo{author}{\bibfnamefont{S.~L.}
  \bibnamefont{{Glashow}}}, \bibinfo{journal}{\nat}
  \textbf{\bibinfo{volume}{326}}, \bibinfo{pages}{476} (\bibinfo{year}{1987}).

\bibitem[{\citenamefont{{Arnett} and {Rosner}}(1987)}]{Arnett1987}
\bibinfo{author}{\bibfnamefont{W.~D.} \bibnamefont{{Arnett}}} \bibnamefont{and}
  \bibinfo{author}{\bibfnamefont{J.~L.} \bibnamefont{{Rosner}}},
  \bibinfo{journal}{Physical Review Letters} \textbf{\bibinfo{volume}{58}},
  \bibinfo{pages}{1906} (\bibinfo{year}{1987}).

\bibitem[{\citenamefont{{Bilenky} et~al.}(2003)\citenamefont{{Bilenky},
  {Giunti}, {Grifols}, and {Mass{\' o}}}}]{Bilenky2003}
\bibinfo{author}{\bibfnamefont{S.~M.} \bibnamefont{{Bilenky}}},
  \bibinfo{author}{\bibfnamefont{C.}~\bibnamefont{{Giunti}}},
  \bibinfo{author}{\bibfnamefont{J.~A.} \bibnamefont{{Grifols}}},
  \bibnamefont{and} \bibinfo{author}{\bibfnamefont{E.}~\bibnamefont{{Mass{\'
  o}}}}, \bibinfo{journal}{Phys. Rep.} \textbf{\bibinfo{volume}{379}},
  \bibinfo{pages}{69} (\bibinfo{year}{2003}).

\bibitem[{\citenamefont{{Hillebrandt} and {Hoflich}}(1989)}]{Hillebrandt1989}
\bibinfo{author}{\bibfnamefont{W.}~\bibnamefont{{Hillebrandt}}}
  \bibnamefont{and}
  \bibinfo{author}{\bibfnamefont{P.}~\bibnamefont{{Hoflich}}},
  \bibinfo{journal}{Reports of Progress in Physics}
  \textbf{\bibinfo{volume}{52}}, \bibinfo{pages}{1421} (\bibinfo{year}{1989}).

\bibitem[{\citenamefont{{Beacom} et~al.}(2000)\citenamefont{{Beacom}, {Boyd},
  and {Mezzacappa}}}]{Beacom2000}
\bibinfo{author}{\bibfnamefont{J.~F.} \bibnamefont{{Beacom}}},
  \bibinfo{author}{\bibfnamefont{R.~N.} \bibnamefont{{Boyd}}},
  \bibnamefont{and}
  \bibinfo{author}{\bibfnamefont{A.}~\bibnamefont{{Mezzacappa}}},
  \bibinfo{journal}{Physical Review Letters} \textbf{\bibinfo{volume}{85}},
  \bibinfo{pages}{3568} (\bibinfo{year}{2000}).

\bibitem[{\citenamefont{{Longo}}(1988)}]{Longo1988}
\bibinfo{author}{\bibfnamefont{M.~J.} \bibnamefont{{Longo}}},
  \bibinfo{journal}{Physical Review Letters} \textbf{\bibinfo{volume}{60}},
  \bibinfo{pages}{173} (\bibinfo{year}{1988}).

\bibitem[{\citenamefont{{Krauss} and {Tremaine}}(1988)}]{Krauss1988}
\bibinfo{author}{\bibfnamefont{L.~M.} \bibnamefont{{Krauss}}} \bibnamefont{and}
  \bibinfo{author}{\bibfnamefont{S.}~\bibnamefont{{Tremaine}}},
  \bibinfo{journal}{Physical Review Letters} \textbf{\bibinfo{volume}{60}},
  \bibinfo{pages}{176} (\bibinfo{year}{1988}).

\bibitem[{\citenamefont{{Barrow} and {Subramanian}}(1987)}]{Barrow1987}
\bibinfo{author}{\bibfnamefont{J.~D.} \bibnamefont{{Barrow}}} \bibnamefont{and}
  \bibinfo{author}{\bibfnamefont{K.}~\bibnamefont{{Subramanian}}},
  \bibinfo{journal}{\nat} \textbf{\bibinfo{volume}{327}}, \bibinfo{pages}{375}
  (\bibinfo{year}{1987}).

\bibitem[{\citenamefont{{Gerver}}(1988)}]{Gerver1988}
\bibinfo{author}{\bibfnamefont{J.~L.} \bibnamefont{{Gerver}}},
  \bibinfo{journal}{Physics Letters A} \textbf{\bibinfo{volume}{127}},
  \bibinfo{pages}{301} (\bibinfo{year}{1988}).

\bibitem[{\citenamefont{{Escribano} et~al.}(2001)\citenamefont{{Escribano},
  {Fr{\` e}re}, {Monderen}, and {Van Elewyck}}}]{Escribano2001}
\bibinfo{author}{\bibfnamefont{R.}~\bibnamefont{{Escribano}}},
  \bibinfo{author}{\bibfnamefont{J.-M.} \bibnamefont{{Fr{\` e}re}}},
  \bibinfo{author}{\bibfnamefont{D.}~\bibnamefont{{Monderen}}},
  \bibnamefont{and} \bibinfo{author}{\bibfnamefont{V.}~\bibnamefont{{Van
  Elewyck}}}, \bibinfo{journal}{Physics Letters B}
  \textbf{\bibinfo{volume}{512}}, \bibinfo{pages}{8} (\bibinfo{year}{2001}).

\bibitem[{\citenamefont{{Brill} and {Wheeler}}(1957)}]{Brill1957}
\bibinfo{author}{\bibfnamefont{D.~R.} \bibnamefont{{Brill}}} \bibnamefont{and}
  \bibinfo{author}{\bibfnamefont{J.~A.} \bibnamefont{{Wheeler}}},
  \bibinfo{journal}{Reviews of Modern Physics} \textbf{\bibinfo{volume}{29}},
  \bibinfo{pages}{465} (\bibinfo{year}{1957}).

\bibitem[{\citenamefont{{Stodolsky}}(1979)}]{Stodolsky1979}
\bibinfo{author}{\bibfnamefont{L.}~\bibnamefont{{Stodolsky}}},
  \bibinfo{journal}{General Relativity and Gravitation}
  \textbf{\bibinfo{volume}{11}}, \bibinfo{pages}{391} (\bibinfo{year}{1979}).

\bibitem[{\citenamefont{{Fornengo} et~al.}(1997)\citenamefont{{Fornengo},
  {Giunti}, {Kim}, and {Song}}}]{Fornengo1997}
\bibinfo{author}{\bibfnamefont{N.}~\bibnamefont{{Fornengo}}},
  \bibinfo{author}{\bibfnamefont{C.}~\bibnamefont{{Giunti}}},
  \bibinfo{author}{\bibfnamefont{C.~W.} \bibnamefont{{Kim}}}, \bibnamefont{and}
  \bibinfo{author}{\bibfnamefont{J.}~\bibnamefont{{Song}}},
  \bibinfo{journal}{Phys. Rev. D} \textbf{\bibinfo{volume}{56}},
  \bibinfo{pages}{1895} (\bibinfo{year}{1997}).

\bibitem[{\citenamefont{{Bhattacharya}
  et~al.}(1999)\citenamefont{{Bhattacharya}, {Habib}, and
  {Mottola}}}]{Bhattacharya1999}
\bibinfo{author}{\bibfnamefont{T.}~\bibnamefont{{Bhattacharya}}},
  \bibinfo{author}{\bibfnamefont{S.}~\bibnamefont{{Habib}}}, \bibnamefont{and}
  \bibinfo{author}{\bibfnamefont{E.}~\bibnamefont{{Mottola}}},
  \bibinfo{journal}{Phys. Rev. D} \textbf{\bibinfo{volume}{59}},
  \bibinfo{pages}{67301} (\bibinfo{year}{1999}).

\bibitem[{\citenamefont{{Kojima}}(1996)}]{Kojima1996}
\bibinfo{author}{\bibfnamefont{Y.}~\bibnamefont{{Kojima}}},
  \bibinfo{journal}{Modern Physics Letters A} \textbf{\bibinfo{volume}{11}},
  \bibinfo{pages}{2965} (\bibinfo{year}{1996}).

\bibitem[{\citenamefont{{Cardall} and {Fuller}}(1997)}]{Cardall1997}
\bibinfo{author}{\bibfnamefont{C.~Y.} \bibnamefont{{Cardall}}}
  \bibnamefont{and} \bibinfo{author}{\bibfnamefont{G.~M.}
  \bibnamefont{{Fuller}}}, \bibinfo{journal}{Phys. Rev. D}
  \textbf{\bibinfo{volume}{55}}, \bibinfo{pages}{7960} (\bibinfo{year}{1997}).

\bibitem[{\citenamefont{{Konno} and {Kasai}}(1998)}]{Konno1998}
\bibinfo{author}{\bibfnamefont{K.}~\bibnamefont{{Konno}}} \bibnamefont{and}
  \bibinfo{author}{\bibfnamefont{M.}~\bibnamefont{{Kasai}}},
  \bibinfo{journal}{Progress of Theoretical Physics}
  \textbf{\bibinfo{volume}{100}}, \bibinfo{pages}{1145} (\bibinfo{year}{1998}).

\bibitem[{\citenamefont{{Alsing} et~al.}(2001)\citenamefont{{Alsing}, {Evans},
  and {Nandi}}}]{Alsing2001}
\bibinfo{author}{\bibfnamefont{P.~M.} \bibnamefont{{Alsing}}},
  \bibinfo{author}{\bibfnamefont{J.~C.} \bibnamefont{{Evans}}},
  \bibnamefont{and} \bibinfo{author}{\bibfnamefont{K.~K.}
  \bibnamefont{{Nandi}}}, \bibinfo{journal}{General Relativity and Gravitation}
  \textbf{\bibinfo{volume}{33}}, \bibinfo{pages}{1459} (\bibinfo{year}{2001}).

\bibitem[{\citenamefont{{Ahluwalia} and {Burgard}}(1996)}]{Ahluwalia1996}
\bibinfo{author}{\bibfnamefont{D.~V.} \bibnamefont{{Ahluwalia}}}
  \bibnamefont{and}
  \bibinfo{author}{\bibfnamefont{C.}~\bibnamefont{{Burgard}}},
  \bibinfo{journal}{General Relativity and Gravitation}
  \textbf{\bibinfo{volume}{28}}, \bibinfo{pages}{1161} (\bibinfo{year}{1996}).

\bibitem[{\citenamefont{{Wudka}}(2001)}]{Wudka2001}
\bibinfo{author}{\bibfnamefont{J.}~\bibnamefont{{Wudka}}},
  \bibinfo{journal}{\prd} \textbf{\bibinfo{volume}{64}}, \bibinfo{pages}{65009}
  (\bibinfo{year}{2001}).

\bibitem[{\citenamefont{{Ahluwalia} and {Burgard}}(1998)}]{Ahluwalia1998}
\bibinfo{author}{\bibfnamefont{D.~V.} \bibnamefont{{Ahluwalia}}}
  \bibnamefont{and}
  \bibinfo{author}{\bibfnamefont{C.}~\bibnamefont{{Burgard}}},
  \bibinfo{journal}{\prd} \textbf{\bibinfo{volume}{57}}, \bibinfo{pages}{4724}
  (\bibinfo{year}{1998}).

\bibitem[{\citenamefont{{Grossman} and {Lipkin}}(1997)}]{Grossman1997}
\bibinfo{author}{\bibfnamefont{Y.}~\bibnamefont{{Grossman}}} \bibnamefont{and}
  \bibinfo{author}{\bibfnamefont{H.~J.} \bibnamefont{{Lipkin}}},
  \bibinfo{journal}{\prd} \textbf{\bibinfo{volume}{55}}, \bibinfo{pages}{2760}
  (\bibinfo{year}{1997}).

\bibitem[{\citenamefont{{Zhang} and {Beesham}}(2001)}]{Zhang2001}
\bibinfo{author}{\bibfnamefont{C.~M.} \bibnamefont{{Zhang}}} \bibnamefont{and}
  \bibinfo{author}{\bibfnamefont{A.}~\bibnamefont{{Beesham}}},
  \bibinfo{journal}{General Relativity and Gravitation}
  \textbf{\bibinfo{volume}{33}}, \bibinfo{pages}{1011} (\bibinfo{year}{2001}).

\bibitem[{\citenamefont{{Linet} and {Teyssandier}}(2002)}]{Linet2002}
\bibinfo{author}{\bibfnamefont{B.}~\bibnamefont{{Linet}}} \bibnamefont{and}
  \bibinfo{author}{\bibfnamefont{P.}~\bibnamefont{{Teyssandier}}},
  \bibinfo{journal}{ArXiv General Relativity and Quantum Cosmology e-prints}
  pp. \bibinfo{pages}{6056--+} (\bibinfo{year}{2002}), \eprint{gr-qc/0206056}.

\bibitem[{\citenamefont{{Zhang} and {Beesham}}(2003)}]{Zhang2003}
\bibinfo{author}{\bibfnamefont{C.~M.} \bibnamefont{{Zhang}}} \bibnamefont{and}
  \bibinfo{author}{\bibfnamefont{A.}~\bibnamefont{{Beesham}}},
  \bibinfo{journal}{International Journal of Modern Physics D}
  \textbf{\bibinfo{volume}{12}}, \bibinfo{pages}{727} (\bibinfo{year}{2003}).

\bibitem[{\citenamefont{{Giunti}}(2002{\natexlab{a}})}]{Giunti2002}
\bibinfo{author}{\bibfnamefont{C.}~\bibnamefont{{Giunti}}},
  \bibinfo{journal}{ArXiv High Energy Physics - Phenomenology e-prints}
  (\bibinfo{year}{2002}{\natexlab{a}}), \eprint{hep-ph/0202063}.

\bibitem[{\citenamefont{{Beuthe}}(2003)}]{Beuthe2001}
\bibinfo{author}{\bibfnamefont{M.}~\bibnamefont{{Beuthe}}},
  \bibinfo{journal}{Phys. Rep.} \textbf{\bibinfo{volume}{375}},
  \bibinfo{pages}{105} (\bibinfo{year}{2003}).

\bibitem[{\citenamefont{{Beuthe}}(2002)}]{Beuthe2002}
\bibinfo{author}{\bibfnamefont{M.}~\bibnamefont{{Beuthe}}},
  \bibinfo{journal}{\prd} \textbf{\bibinfo{volume}{66}}, \bibinfo{pages}{13003}
  (\bibinfo{year}{2002}).

\bibitem[{\citenamefont{{Giunti}}(2003)}]{Giunti2003}
\bibinfo{author}{\bibfnamefont{C.}~\bibnamefont{{Giunti}}},
  \bibinfo{journal}{ArXiv High Energy Physics - Phenomenology e-prints}
  (\bibinfo{year}{2003}), \eprint{hep-ph/0302026}.

\bibitem[{\citenamefont{{Giunti} et~al.}(1991)\citenamefont{{Giunti}, {Kim},
  and {Lee}}}]{Giunti1991}
\bibinfo{author}{\bibfnamefont{C.}~\bibnamefont{{Giunti}}},
  \bibinfo{author}{\bibfnamefont{C.~W.} \bibnamefont{{Kim}}}, \bibnamefont{and}
  \bibinfo{author}{\bibfnamefont{U.~W.} \bibnamefont{{Lee}}},
  \bibinfo{journal}{Phys. Rev. D} \textbf{\bibinfo{volume}{44}},
  \bibinfo{pages}{3635} (\bibinfo{year}{1991}).

\bibitem[{\citenamefont{{Misner} et~al.}(1973)\citenamefont{{Misner}, {Thorne},
  and {Wheeler}}}]{Misner1973}
\bibinfo{author}{\bibfnamefont{C.~W.} \bibnamefont{{Misner}}},
  \bibinfo{author}{\bibfnamefont{K.~S.} \bibnamefont{{Thorne}}},
  \bibnamefont{and} \bibinfo{author}{\bibfnamefont{J.~A.}
  \bibnamefont{{Wheeler}}}, \emph{\bibinfo{title}{{Gravitation}}}
  (\bibinfo{publisher}{San Francisco: W.H.~Freeman and Co., 1973},
  \bibinfo{year}{1973}).

\bibitem[{\citenamefont{{Giunti}}(2001)}]{Giunti2001}
\bibinfo{author}{\bibfnamefont{C.}~\bibnamefont{{Giunti}}},
  \bibinfo{journal}{Modern Physics Letters A} \textbf{\bibinfo{volume}{16}},
  \bibinfo{pages}{2363} (\bibinfo{year}{2001}).

\bibitem[{\citenamefont{{Giunti}}(2002{\natexlab{b}})}]{Giunti2002d}
\bibinfo{author}{\bibfnamefont{C.}~\bibnamefont{{Giunti}}},
  \bibinfo{journal}{ArXiv High Energy Physics - Phenomenology e-prints}
  (\bibinfo{year}{2002}{\natexlab{b}}), \eprint{hep-ph/0205014}.

\bibitem[{\citenamefont{{Bhattacharya}
  et~al.}(1996)\citenamefont{{Bhattacharya}, {Habib}, and
  {Mottola}}}]{Bhattacharya1996}
\bibinfo{author}{\bibfnamefont{T.}~\bibnamefont{{Bhattacharya}}},
  \bibinfo{author}{\bibfnamefont{S.}~\bibnamefont{{Habib}}}, \bibnamefont{and}
  \bibinfo{author}{\bibfnamefont{E.}~\bibnamefont{{Mottola}}},
  \bibinfo{journal}{pre-print(gr-qc/9605074)}  (\bibinfo{year}{1996}).

\bibitem[{\citenamefont{{Landau} and {Lifshitz}}(1975)}]{Landau1975}
\bibinfo{author}{\bibfnamefont{L.~D.} \bibnamefont{{Landau}}} \bibnamefont{and}
  \bibinfo{author}{\bibfnamefont{E.~M.} \bibnamefont{{Lifshitz}}},
  \emph{\bibinfo{title}{{The classical theory of fields}}}
  (\bibinfo{publisher}{Course of theoretical physics - Pergamon International
  Library of Science, Technology, Engineering and Social Studies, Oxford:
  Pergamon Press, 1975, 4th rev.engl.ed.}, \bibinfo{year}{1975}).

\bibitem[{\citenamefont{{Weinberg}}(1972)}]{Weinberg1972}
\bibinfo{author}{\bibfnamefont{S.}~\bibnamefont{{Weinberg}}},
  \emph{\bibinfo{title}{{Gravitation and cosmology: Principles and applications
  of the general theory of relativity}}} (\bibinfo{publisher}{New York: Wiley,
  |c1972}, \bibinfo{year}{1972}).

\bibitem[{\citenamefont{{Evans} et~al.}(1996)\citenamefont{{Evans}, {Nandi},
  and {Islam}}}]{Evans1996}
\bibinfo{author}{\bibfnamefont{J.}~\bibnamefont{{Evans}}},
  \bibinfo{author}{\bibfnamefont{K.~K.} \bibnamefont{{Nandi}}},
  \bibnamefont{and} \bibinfo{author}{\bibfnamefont{A.}~\bibnamefont{{Islam}}},
  \bibinfo{journal}{American Journal of Physics} \textbf{\bibinfo{volume}{64}},
  \bibinfo{pages}{1404} (\bibinfo{year}{1996}).

\bibitem[{\citenamefont{{Schneider} et~al.}(1992)\citenamefont{{Schneider},
  {Ehlers}, and {Falco}}}]{Schneider1992}
\bibinfo{author}{\bibfnamefont{P.}~\bibnamefont{{Schneider}}},
  \bibinfo{author}{\bibfnamefont{J.}~\bibnamefont{{Ehlers}}}, \bibnamefont{and}
  \bibinfo{author}{\bibfnamefont{E.~E.} \bibnamefont{{Falco}}},
  \emph{\bibinfo{title}{{Gravitational Lenses}}}
  (\bibinfo{publisher}{Gravitational Lenses, XIV, 560 pp.~112
  figs..~Springer-Verlag Berlin Heidelberg New York.~ Also Astronomy and
  Astrophysics Library}, \bibinfo{year}{1992}).

\bibitem[{\citenamefont{{Nakaya} and {al.}}(2002{\natexlab{a}})}]{Nakaya2002}
\bibinfo{author}{\bibfnamefont{T.}~\bibnamefont{{Nakaya}}} \bibnamefont{and}
  \bibinfo{author}{\bibnamefont{{al.}}}
  (\bibinfo{collaboration}{SUPER-KAMIOKANDE}), \bibinfo{journal}{ArXiv High
  Energy Physics - Experimental e-prints}
  (\bibinfo{year}{2002}{\natexlab{a}}), \eprint{hep-ex/0209036}.

\bibitem[{\citenamefont{{Crocker} and {Mortlock}}(2003)}]{Crocker2003}
\bibinfo{author}{\bibfnamefont{R.~M.} \bibnamefont{{Crocker}}}
  \bibnamefont{and}
  \bibinfo{author}{\bibfnamefont{D.}~\bibnamefont{{Mortlock}}},
  \bibinfo{journal}{{forthcoming}}  (\bibinfo{year}{2003}).

\bibitem[{\citenamefont{{Kayser}}(1981)}]{Kayser1981}
\bibinfo{author}{\bibfnamefont{B.}~\bibnamefont{{Kayser}}},
  \bibinfo{journal}{\prd} \textbf{\bibinfo{volume}{24}}, \bibinfo{pages}{110}
  (\bibinfo{year}{1981}).

\bibitem[{\citenamefont{{Kim} and {Pevsner}}(1993)}]{Kim1993}
\bibinfo{author}{\bibfnamefont{C.~W.} \bibnamefont{{Kim}}} \bibnamefont{and}
  \bibinfo{author}{\bibfnamefont{A.}~\bibnamefont{{Pevsner}}},
  \emph{\bibinfo{title}{{Neutrinos in Physics and Astrophysics (Contemporary
  Concepts in Physics, 8)}}} (\bibinfo{publisher}{Chur, Switzerland: Harwood,
  1993}, \bibinfo{year}{1993}).

\bibitem[{\citenamefont{{Nussinov}}(1976)}]{Nussinov1976}
\bibinfo{author}{\bibfnamefont{S.}~\bibnamefont{{Nussinov}}},
  \bibinfo{journal}{Physics Letters B} \textbf{\bibinfo{volume}{63}},
  \bibinfo{pages}{201} (\bibinfo{year}{1976}).

\bibitem[{\citenamefont{{Anada} and {Nishimura}}(1988)}]{Anada1988}
\bibinfo{author}{\bibfnamefont{H.}~\bibnamefont{{Anada}}} \bibnamefont{and}
  \bibinfo{author}{\bibfnamefont{H.}~\bibnamefont{{Nishimura}}},
  \bibinfo{journal}{\prd} \textbf{\bibinfo{volume}{37}}, \bibinfo{pages}{552}
  (\bibinfo{year}{1988}).

\bibitem[{\citenamefont{{Beacom} and {Vogel}}(1999)}]{Beacom1999}
\bibinfo{author}{\bibfnamefont{J.~F.} \bibnamefont{{Beacom}}} \bibnamefont{and}
  \bibinfo{author}{\bibfnamefont{P.}~\bibnamefont{{Vogel}}},
  \bibinfo{journal}{Phys. Rev. D} \textbf{\bibinfo{volume}{60}},
  \bibinfo{pages}{33007} (\bibinfo{year}{1999}).

\bibitem[{\citenamefont{{Nakaya} and {al.}}(2002{\natexlab{b}})}]{Nakaya2002a}
\bibinfo{author}{\bibfnamefont{T.}~\bibnamefont{{Nakaya}}} \bibnamefont{and}
  \bibinfo{author}{\bibnamefont{{al.}}}
  (\bibinfo{collaboration}{SUPER-KAMIOKANDE}), \bibinfo{journal}{presented at
  the XXth Int. Conf. on Neutrino Physics and Astrophysics, Munich}
  (\bibinfo{year}{2002}{\natexlab{b}}).

\bibitem[{\citenamefont{{Antonioli} and {al.}}(1999)}]{Antonioli1999}
\bibinfo{author}{\bibfnamefont{P.}~\bibnamefont{{Antonioli}}} \bibnamefont{and}
  \bibinfo{author}{\bibnamefont{{al.}}}, \bibinfo{journal}{Nuclear Instruments
  and Methods in Physics Research A} \textbf{\bibinfo{volume}{433}},
  \bibinfo{pages}{104} (\bibinfo{year}{1999}).

\bibitem[{\citenamefont{{Jung}}(1999)}]{Jung2000}
\bibinfo{author}{\bibfnamefont{C.~K.} \bibnamefont{{Jung}}},
  \bibinfo{journal}{ArXiv High Energy Physics - Experimental e-prints}
  (\bibinfo{year}{1999}), \eprint{pre-print(hep-ex/0005046}.

\bibitem[{\citenamefont{{Anada} and {Nishimura}}(1990)}]{Anada1990}
\bibinfo{author}{\bibfnamefont{H.}~\bibnamefont{{Anada}}} \bibnamefont{and}
  \bibinfo{author}{\bibfnamefont{H.}~\bibnamefont{{Nishimura}}},
  \bibinfo{journal}{\prd} \textbf{\bibinfo{volume}{41}}, \bibinfo{pages}{2379}
  (\bibinfo{year}{1990}).

\bibitem[{\citenamefont{{Sakamoto} et~al.}(2003)\citenamefont{{Sakamoto},
  {Chiba}, and {Beers}}}]{Sakamoto2003}
\bibinfo{author}{\bibfnamefont{T.}~\bibnamefont{{Sakamoto}}},
  \bibinfo{author}{\bibfnamefont{M.}~\bibnamefont{{Chiba}}}, \bibnamefont{and}
  \bibinfo{author}{\bibfnamefont{T.~C.} \bibnamefont{{Beers}}},
  \textbf{\bibinfo{volume}{397}}, \bibinfo{pages}{899} (\bibinfo{year}{2003}).

\bibitem[{\citenamefont{{Paczynski}}(1986)}]{Paczynski1986}
\bibinfo{author}{\bibfnamefont{B.}~\bibnamefont{{Paczynski}}},
  \bibinfo{journal}{ApJ} \textbf{\bibinfo{volume}{304}}, \bibinfo{pages}{1}
  (\bibinfo{year}{1986}).

\bibitem[{\citenamefont{{Alcock} et~al.}(2000)\citenamefont{{Alcock},
  {Allsman}, {Alves}, {Axelrod}, {Becker}, {Bennett}, {Cook}, {Drake},
  {Freeman}, {Geha} et~al.}}]{Alcock2000}
\bibinfo{author}{\bibfnamefont{C.}~\bibnamefont{{Alcock}}},
  \bibinfo{author}{\bibfnamefont{R.~A.} \bibnamefont{{Allsman}}},
  \bibinfo{author}{\bibfnamefont{D.~R.} \bibnamefont{{Alves}}},
  \bibinfo{author}{\bibfnamefont{T.~S.} \bibnamefont{{Axelrod}}},
  \bibinfo{author}{\bibfnamefont{A.~C.} \bibnamefont{{Becker}}},
  \bibinfo{author}{\bibfnamefont{D.~P.} \bibnamefont{{Bennett}}},
  \bibinfo{author}{\bibfnamefont{K.~H.} \bibnamefont{{Cook}}},
  \bibinfo{author}{\bibfnamefont{A.~J.} \bibnamefont{{Drake}}},
  \bibinfo{author}{\bibfnamefont{K.~C.} \bibnamefont{{Freeman}}},
  \bibinfo{author}{\bibfnamefont{M.}~\bibnamefont{{Geha}}},
  \bibnamefont{et~al.}, \bibinfo{journal}{ApJ} \textbf{\bibinfo{volume}{541}},
  \bibinfo{pages}{734} (\bibinfo{year}{2000}).

\bibitem[{\citenamefont{{Wozniak} et~al.}(2001)\citenamefont{{Wozniak},
  {Udalski}, {Szymanski}, {Kubiak}, {Pietrzynski}, {Soszynski}, and
  {Zebrun}}}]{Wozniak2001}
\bibinfo{author}{\bibfnamefont{P.~R.} \bibnamefont{{Wozniak}}},
  \bibinfo{author}{\bibfnamefont{A.}~\bibnamefont{{Udalski}}},
  \bibinfo{author}{\bibfnamefont{M.}~\bibnamefont{{Szymanski}}},
  \bibinfo{author}{\bibfnamefont{M.}~\bibnamefont{{Kubiak}}},
  \bibinfo{author}{\bibfnamefont{G.}~\bibnamefont{{Pietrzynski}}},
  \bibinfo{author}{\bibfnamefont{I.}~\bibnamefont{{Soszynski}}},
  \bibnamefont{and} \bibinfo{author}{\bibfnamefont{K.}~\bibnamefont{{Zebrun}}},
  \bibinfo{journal}{Acta Astronomica} \textbf{\bibinfo{volume}{51}},
  \bibinfo{pages}{175} (\bibinfo{year}{2001}).

\bibitem[{\citenamefont{{Wagner} and {Weiler}}(1997)}]{Wagner1998}
\bibinfo{author}{\bibfnamefont{D.~J.} \bibnamefont{{Wagner}}} \bibnamefont{and}
  \bibinfo{author}{\bibfnamefont{T.~J.} \bibnamefont{{Weiler}}},
  \bibinfo{journal}{Modern Physics Letters A} \textbf{\bibinfo{volume}{12}},
  \bibinfo{pages}{2497} (\bibinfo{year}{1997}).

\bibitem[{\citenamefont{{Mbonye}}(2002)}]{Mbonye2001}
\bibinfo{author}{\bibfnamefont{M.~R.} \bibnamefont{{Mbonye}}},
  \bibinfo{journal}{{Gen. Rel. Grav.}} \textbf{\bibinfo{volume}{34}},
  \bibinfo{pages}{1865} (\bibinfo{year}{2002}).

\bibitem[{\citenamefont{{Ragazzoni} et~al.}(2003)\citenamefont{{Ragazzoni},
  {Turatto}, and {Gaessler}}}]{Ragazzoni2003}
\bibinfo{author}{\bibfnamefont{R.}~\bibnamefont{{Ragazzoni}}},
  \bibinfo{author}{\bibfnamefont{M.}~\bibnamefont{{Turatto}}},
  \bibnamefont{and}
  \bibinfo{author}{\bibfnamefont{W.}~\bibnamefont{{Gaessler}}},
  \bibinfo{journal}{ApJL} \textbf{\bibinfo{volume}{587}}, \bibinfo{pages}{L1}
  (\bibinfo{year}{2003}).

\bibitem[{\citenamefont{{Bolton} and {Burles}}(2003)}]{Bolton2003}
\bibinfo{author}{\bibfnamefont{A.~S.} \bibnamefont{{Bolton}}} \bibnamefont{and}
  \bibinfo{author}{\bibfnamefont{S.}~\bibnamefont{{Burles}}},
  \bibinfo{journal}{ApJ} \textbf{\bibinfo{volume}{592}}, \bibinfo{pages}{17}
  (\bibinfo{year}{2003}).

\bibitem[{\citenamefont{{Refsdal}}(1964)}]{Refsdal1964}
\bibinfo{author}{\bibfnamefont{S.}~\bibnamefont{{Refsdal}}},
  \bibinfo{journal}{MNRAS} \textbf{\bibinfo{volume}{128}}, \bibinfo{pages}{307}
  (\bibinfo{year}{1964}).

\bibitem[{\citenamefont{{Giunti} and {Kim}}(1998)}]{Giunti1998a}
\bibinfo{author}{\bibfnamefont{C.}~\bibnamefont{{Giunti}}} \bibnamefont{and}
  \bibinfo{author}{\bibfnamefont{C.~W.} \bibnamefont{{Kim}}},
  \bibinfo{journal}{Phys. Rev. D} \textbf{\bibinfo{volume}{58}},
  \bibinfo{pages}{17301} (\bibinfo{year}{1998}).

\end{thebibliography}
\newpage
\section{Appendix: Wave Packet Treatment of Neutrino Beam Splitter}

In this appendix we set forth a full (Gaussian) wave packet
treatment of the neutrino beam spliter {\it Gedanken Experiment} (treated in
terms of plane waves in \S 3).
Note that the results we derive for the exponential damping
terms (in the equation for the oscillation probability analog --
see Eq.~(\ref{eqn_app_bs''''}))
serve as an heuristic justification of the
treatment of decoherence we present in \S 6.

We write the ket associated with the 
neutrino flavour eigenstate $\alpha$ that has propagated from
 the source spacetime position $A = (x_A, t_A)$
to
detection position 
$B = (x_B, t_B)$ as
\begin{multline}
\label{eqn_app_bs}
\vert \nu_\alpha,B \rangle
= N \sum_p \sqrt{I_p} \sum_j U_{\alpha j} \\
\times
\int \mathrm{d}E \exp[-\mathrm{i} \Phi^p_j(E; L^{AB}_p, T^{AB})] A_j(E) 
\vert \nu_j \rangle,
\end{multline}
where the various quantities are as explained in \S 3.
The amplitude for a neutrino created as type $\alpha$ to be
detected as type $\beta$ at the spacetime position, $B$, 
of the detection event is then:
\begin{multline}
\label{eqn_app_bs'}
\langle \nu_\beta \vert \nu_\alpha; A,B \rangle 
= N \sum_p \sqrt{I_p} \sum_j U_{\alpha j} U_{\beta j}^* \\
 \times  \int \mathrm{d}E  A_j(E) \exp[-\mathrm{i} \Phi^p_j(E; L^{AB}_p, T^{AB})].
\end{multline}
Again, we can get rid of the unwanted dependence on time by averaging over $T^{AB}$
in the above to determine
a time-averaged oscillation probability {\it analog} 
at the detector position $x_B$ 
\cite{Beuthe2001,Beuthe2002,Giunti2003}.
This  gives us that
\begin{multline}
\label{eqn_app_bs''}
\vert \langle \nu_\beta \vert \nu_\alpha; x_A,x_B \rangle \vert^2
\propto \int \mathrm{d}T^{AB} 
\vert \langle \nu_\beta \vert \nu_\alpha; A,B \rangle \vert^2 \\
\propto \vert N \vert^2 \sum_{pq} \sqrt{I_pI_q} \sum_{jk} 
U_{\alpha j} U_{\beta j}^* U_{\beta k} U_{\alpha k}^*  \\
\times \int \mathrm{d}E A_j(E) A_k^*(E)
\exp[\mathrm{i}(p_j(E)L^{AB}_p - p_k(E)L^{AB}_q)],
\end{multline}
where one integral over energy has disappeared because
of the $\delta(E - E')$ that arises from the integration over time.

Assuming a Gaussian form for the wavepackets leads to
\be
A_j \propto \exp \left[-\frac{(E - {\bar E}_j)^2}{4 (\sigma^{(j)}_E)^2} \right]
\simeq \exp \left[-\frac{(E - {\bar E}_j)^2}{4 \sigma^2_E} \right],
\ee
where ${\bar E}_j$ 
is the
peak or average energy of \MES \ $j$ and
we employ the very good approximation \citep{Giunti1991} that 
the wavepacket spread is the same for different \MES s. This gives us that
\begin{multline}
A_j(E) A_k^*(E) \\
\propto \exp\left[-\frac{2E^2 - 2E({\bar E}_j + {\bar E}_k) + 
({\bar E}^2_j + {\bar E}^2_k)}
{4 \sigma_E^2} \right].
\end{multline}
Now defining the peak momentum of \MES\ $j$ via ${\bar p}_j = 
({\bar E}_j^2 - m_j^2)^{1/2}$, 
we can write
\be
\label{eqn_app_mtmdiff}
p_j(E) \equiv {\bar p}_j + \Delta p_j(E),
\ee
where
\bea
\label{eqn_app_mtmdiff'}
&\Delta p_j(E)
&\equiv p_j(E) - {\bar p}_j \nn \\
&&= \sqrt{E^2 - m_j^2} - \sqrt{{\bar E}_j^2 - m_j^2} \nn \\
&&\simeq E - \frac{m_j^2}{2E} - {\bar E}_j + \frac{m_j^2}{2{\bar E}_j} \nn \\
&&= (E - {\bar E}_j)\left(1 - \frac{m_j^2}{2E{\bar E}_j}\right).
\eea
Employing the group velocity of \MES \ $j$, viz. 
\be
\label{eqn_app_mtmdiff''}
v_j = {\bar p}_j/{\bar E}_j = \frac{\sqrt{{\bar E}^2 - m_j^2}}{{\bar E}_j} 
\simeq 1 - \frac{m_j^2}{2 {\bar E}_j^2},
\ee
we determine that $\Delta p_j(E) \simeq (E - {\bar E}_j)v_j$ \ \footnote{Note 
that
here we are ignoring
the energy dependence of the velocities of the different plane
wave components of each wavepacket so that our calculation makes no
account of dispersive effects which, in general, extend the coherence length:
see \cite{Kim1993}.}, so that
\be
\label{eqn_app_Beuthe''}
p_j(E) \simeq {\bar p}_j + (E - {\bar E}_j)v_j
\ee
We find, then, after a simple calculation that the oscillation probability 
analog becomes
\begin{multline}
\label{eqn_app_bs'''}
\vert \langle \nu_\beta \vert \nu_\alpha; x_A,x_B \rangle \vert^2\\
\simeq \frac{1}{\sum_{rs} \sqrt{I_rI_s}} \sum_{pq} \sqrt{I_pI_q} \sum_{jk} 
U_{\alpha j} U_{\beta j}^* U_{\beta k} U_{\alpha k}^*  \\
\times  \exp\Bigl\{-\mathrm{i}\Bigl[ (v_jL^{AB}_p + v_kL^{AB}_q)
  \frac{{\bar E}_j - {\bar E}_k}{2}
 \\
\shoveright{- (p_jL^{AB}_p - p_kL^{AB}_q)\Bigr] \Bigr\} } \\
 \times \exp\left[-\sigma_x^2\frac{({\bar E}_j - {\bar E}_k)^2}{2}\right]
\exp\left[-\frac{(v_jL^{AB}_p - v_kL^{AB}_q)^2}{8\sigma_x^2}\right]. \\
\end{multline}
Note that in the above calculation, though we have been employing the 
group velocity,
this does not -- and should not be seen to -- 
enter into the phase in any fundamental way \citep{Giunti2002}. 
Indeed, the phase can
be calculated entirely with plane waves (see \S 3) and, therefore, totally
without reference to wavepacket notions like group velocity -- though the
exponential damping factors above critically depend on these. 
Note also that, again, the normalization has been determined by requiring
that 
$\vert \langle \nu_\beta \vert \nu_\alpha; x_A,x_B \rangle \vert^2 \leq 1$.

The two damping factors in Eq.~(\ref{eqn_bs'''}) can be traced back
to considerations following from (i) source localization and
the (ii) requirement  for overlap of wavepackets at the detector's position.
Observe that the second damping term accounts for an interesting possibility:
having  
a heavier -- and slower -- \MES \ travels down
the shorter path and the lighter \MES \ down the longer path,  will tend to
restore coherence.

As a further particularization  of the 
expression for neutrino oscillation `probability' with 
an imaginary beamsplitter, we take the expressions for 
the energies, momenta, and velocities of the various wavefunctions
given in terms of expansions around the energy in massless limit, $E_0$, viz:
\begin{eqnarray} 
\label{app_xieqn}
& {\bar E}_i & \simeq \ E_0 \ + \ (1-\xi)\frac{m_i^2}{2 E_0} \\
& {\bar p}_i & \simeq \ E_0 \ - \ \xi\frac{m_i^2}{2 E_0},
\end{eqnarray}
where $\xi$ is a dimensionless parameter of order unity
 determined by kinematical
considerations \citep{Giunti2002}. 
The group velocity will then be
\begin{eqnarray}
\label{eqn_app_vel}
& v_i & = \frac{{\bar p}_i}{{\bar E}_i} \nonumber \\
&& \simeq 1 - \frac{m_i^2}{2 E_0^2} \;.
\end{eqnarray}
Employing the above we find
\begin{multline}
\label{eqn_app_bs''''}
\vert \langle \nu_\beta \vert \nu_\alpha; x_A,x_B \rangle \vert^2\\
\simeq \frac{1}{\sum_{rs} \sqrt{I_rI_s}} \sum_{pq} \sqrt{I_pI_q} \sum_{jk} 
U_{\alpha j} U_{\beta j}^* U_{\beta k} U_{\alpha k}^*  \\
\times  \exp\left[-\mathrm{i}\Delta \Phi_{jk}^{pq}\right] \\
\times \exp\left[-\frac{\sigma_x^2}{2}\left(\xi\frac{\delta m^2_{jk}}{2 E_0}
\right)^2\right]\\
\times \exp\left[-\frac{1}{2}\left(\frac{L^{AB}_p - L^{AB}_q}{2 \sigma_x} 
- \frac{1}{2 E_0^2}\frac{m_j^2 L^{AB}_p - m_k^2 L^{AB}_q}{2 \sigma_x}\right)^2\right],
\end{multline}
where the phase difference is given by
\begin{multline}
\label{eqn_app_splitterphase}
\Delta \Phi_{jk}^{pq} \equiv 
- \left(E_0 + \xi\frac{m_j^2 + m_k^2}{4E_0}\right)(L^{AB}_p - L^{AB}_q) \\
+ \left(\frac{m_j^2 L^{AB}_p - m_k^2 L^{AB}_q}{2 E_0}\right).
\end{multline}
Note that we can take the  plane wave limit of the
above equation
by setting
$\xi \to 0$ and $\sigma_x \to \infty$. This allows us to recover Eq.~(\ref{eqn_bs'''}).

\end{document}